\pdfoutput=1

\documentclass[twocolumn,twocolappendix]{aastex701}

\shorttitle{SLACS Spatially Resolved Kinematics II}
\shortauthors{Knabel et al.}

\graphicspath{{./}}
\usepackage{newtxtext,newtxmath}
\usepackage{amsmath}
\usepackage{lineno}

\newcommand{\moment}[1]{\langle{#1}\rangle}

\newcommand{\bigkappa}{\resizebox{0.2cm}{!}{$\kappa$}}

\begin{document}

\title{Spatially Resolved Kinematics of SLACS Lens Galaxies. II: Breaking Degeneracies with Lensing and Dynamical Models}

\author[0000-0001-5110-6241]{Shawn Knabel}
\affiliation{Department of Physics and Astronomy, University of California, \\
Los Angeles, CA 90095, USA}
\email{shawnknabel@astro.ucla.edu}

\author[0000-0002-8460-0390]{Tommaso Treu}
\affiliation{Department of Physics and Astronomy, University of California, \\
Los Angeles, CA 90095, USA}
\email{tommaso.treu@astro.ucla.edu}

\author[0000-0002-1283-8420]{Michele Cappellari}
\affiliation{Sub-Department of Astrophysics, Department of Physics, University of Oxford, \\
Denys Wilkinson Building, Keble Road, Oxford, OX1 3RH, UK}
\email{michele.cappellari@physics.ox.ac.uk}

\author[0000-0003-3195-5507]{Simon Birrer}
\affiliation{Department of Physics and Astronomy, Stony Brook University, Stony Brook, NY 11794, USA}
\email{s.birrer@stonybrook.edu}

\author[0000-0001-7113-0599]{Xiang-Yu Huang}
\affiliation{Department of Physics and Astronomy, Stony Brook University, Stony Brook, NY 11794, USA}
\email{s.birrer@stonybrook.edu}

\author[0000-0002-5558-888X]{Anowar J. Shajib}
\affiliation{Department of Astronomy \& Astrophysics, University of Chicago, \\
Chicago, IL 60637, USA} 
\affiliation{Kavli Institute for Cosmological Physics, University of Chicago, \\
Chicago, IL 60637, USA}
\altaffiliation{NFHP Einstein Fellow}
\email{ashajib@uchicago.edu}
\altaffiliation{Center for Astronomy, Space Science and Astrophysics, Independent University, \\
Bangladesh, Dhaka 1229, Bangladesh}

\author[0000-0003-1889-0227]{William Sheu}
\affiliation{Department of Physics and Astronomy, University of California, \\
Los Angeles, CA 90095, USA}
\email{william.sheu@astro.ucla.edu}

\correspondingauthor{Shawn Knabel}
\email{shawnknabel@astro.ucla.edu}

\begin{abstract}

We model the dynamical mass density profiles of 14 strong gravitational lens galaxies from the Sloan Lens ACS (SLACS) sample using spatially resolved kinematics obtained from Keck KCWI integral-field spectroscopy. We use the Jeans Anisotropic Modeling (JAM) method, combining 2D kinematic maps with joint constraints from lens models from Hubble Space Telescope imaging. We use informative priors on the anisotropy and intrinsic shape from local galaxies to help break the residual mass-anisotropy degeneracy (MAD). 
We find nearly isothermal power-law total mass density slopes ($\rho_{\rm tot}\propto r^{-\gamma}$) for the sample with a mean of $\gamma = 2.04\pm0.02$ with intrinsic scatter of $0.08^{+0.03}_{-0.02}$. We fit explicitly for deviations from the pure power-law form that are fully sensitive to the mass-sheet degeneracy (MSD) and constrain the value of the mass-sheet parameter $\rm \lambda_{int}$ for each individual galaxy to an average precision of 5.8\%. The mean value of $\rm \lambda_{int}$ for the sample is $1.01\pm0.03$, with intrinsic scatter of $0.11\pm0.03$. Values of $\rm \lambda_{int}$ for individual objects and the scatter in the sample are consistent to $1\sigma$ uncertainty with those found by the Time-Delay COSMOgraphy collaboration's 2025 milestone analysis, which used a spherical analysis of the same dataset, but azimuthally averaged. We thus conclude that on average power-law mass profiles are a good first-order description of the SLACS sample and do not introduce measureable bias in time-delay cosmography. However, our analysis indicates that more flexible mass models should be able to reproduce the highly detailed kinematic datasets more accurately. 

\end{abstract}



\section{Introduction} \label{sec:intro}

The Hubble constant $H_\mathrm{0}$ is currently one of the most debated numbers in cosmology. Measurements from the local Universe and inferences from early-Universe probes under the assumption of $\Lambda$ cold dark matter ($\Lambda$CDM) cosmology have begun to disagree at 5-$\sigma$ significance \citep{verder19_hubble_tension, abdalla_22_hubble_tension, divalentino25_hubble_tension}. If this tension is real, then it could point to the need for new physics beyond the standard $\Lambda$CDM model. Theoretical work has proposed solutions such as a form of early dark energy \citep[e.g.,][]{knox_millea_20, abdalla_22_hubble_tension} and sterile neutrinos \citep[e.g.,][]{pan24_neutrino, escudero26_neutrino}, among many others. For observers, this has led to the prioritization of alternative methods for measuring $H_\mathrm{0}$ and thorough investigation of unknown systematic uncertainties \citep{riess20, riess22, freedman21}.

Time-delay cosmography of lensed quasars \citep{refsdal64} enables the measurement of $H_\mathrm{0}$ with a single step; for an up-to-date review, see \cite{birrer22b}, \cite{treu22}; for a historical perspective,
see \cite{treu_marshall16_tdcosmography}. 
The Time-Delay COSMOgraphy (TDCOSMO) collaboration has analyzed eight time-delay lenses to measure the Hubble constant within 4.6\% error, $H_\mathrm{0} = 74.3^{+3.1}_{-3.7}$ $\mathrm{km}$ $\mathrm{s}^{-1}$ $\mathrm{Mpc}^{-1}$ \citep[][henceforth TDC-25]{tdcosmo25_milestone}. TDC-25 achieved its precision as the culmination of several observational and methodological studies to understand sources of systematic errors and biases \citep[e.g.][]{wells24_tdcxv, knabel_mozumdar25_tdcxix, huang25_triaxiality, mozumdar_knabel26_tdcxxi}.  

TDCOSMO tested several potential sources of uncertainty and concluded that the ``mass-sheet degeneracy" (MSD) is the most significant source of uncertainty in the measurement of $H_\mathrm{0}$ with time-delay cosmography of lensed quasars \citep{birrer20_tdcosmo_iv, millon20b, gilman20, vandevyvere22a, shajib22, vandevyvere22b}. The MSD is unique to lens modeling methods and refers to the non-uniqueness of lens model solutions that can result in the observed lensed image features \citep{falco85, schneider_sluse13, schneider_sluse14}. It is also known as the ``mass-sheet transform" (MST) because it is described by a mathematical transformation of the lens equation that leaves the position and brightness of lensed images unchanged: 

\begin{equation}
    \mathrm{
    \lambda\boldsymbol{\beta} = \boldsymbol{\theta} - \lambda\boldsymbol{\alpha}\boldsymbol{(\theta)} - (1-\lambda)\boldsymbol{\theta}
    }
\end{equation}

\noindent where \textbf{$\beta$} is the true angular position of the background source, \textbf{$\theta$} is the observed angular position of the background source, \textbf{$\alpha$} is the deflection angle, and $\lambda$ is a parameter close to unity that encodes the effect of the mass sheet. In terms of the lensing convergence (dimensionless surface mass density) profile of the deflector galaxy $\kappa \left(\theta\right)$, the transformation is 

\begin{equation}
    \mathrm{
    {\kappa_\mathrm{MST}}({\theta}) = \lambda \times \kappa({\theta}) + (1-\lambda)  
    }
\end{equation}

\noindent describing the presence of an unseen infinite ``sheet" of convergence $\kappa = 1 - \lambda$, and the convergence profile is rescaled by this factor $\lambda$ to leave the lensed images unchanged. While this kind of infinite sheet is unphysical, there are astrophysical sources that can approximate the effect \citep{birrer20_tdcosmo_iv,Blum20}. In time-delay cosmography, the resulting time delays and therefore $H_\mathrm {0}$ are scaled by $\lambda^{-1}$. 

To rigorously build a methodology that breaks these degeneracies and enables a highly precise measurement of $H_0$, it is necessary to go step-by-step through the logical sequence of combining stellar dynamics, gravitational lensing, and time delays. Each step builds upon the previous to improve the accuracy of the mass profile measurements, alleviating the sensitivity of our inferences to biases and improving the error budget.

\begin{enumerate}
    \item \textbf{Lensing constraints:} Gravitational lensing allows one to measure very accurately the enclosed mass within a region approximately given by the Einstein radius \citep{Treu2010ARAA}. Furthermore, with sufficiently high-quality imaging data as those acquired for the time-delay lenses by the TDCOSMO collaboration, one can measure the slope of the mass density profile in the region of the multiple images or lensed arcs \citep[e.g.][]{suyu10}.

    \item \textbf{The Mass-Sheet Degeneracy (MSD):} However, the discussion in the previous step ignores the MSD. Because of the MSD, one can add a sheet of mass density along the line of sight without affecting the image configuration. Thus additional assumptions or external (non-lensing) information are needed to break the MSD. One possible assumption is that of a specific parametric total mass density profile, such as a power law ($\rho_{\rm tot}\propto r^{-\gamma}$). \citep{millon20b} reached 2\% precision, $H_\mathrm{0} = 74.2\pm1.6$ $\mathrm{km}$ $\mathrm{s}^{-1}$ $\mathrm{Mpc}^{-1}$, assuming power-law and composite stellar plus dark matter halo mass density profiles (i.e., $\lambda_\mathrm{int}=1$). In this case, if the true density profile deviates from the parametric form, the assumption may bias the measurement of the mass profile and, consequently, of $H_0$ \citep{birrer20_tdcosmo_iv}. External information includes stellar kinematics \citep{birrer_treu21} or knowledge of the absolute brightness or size of the source \citep{Holz2001, birrer_dhawan_shajib22_sn}.
    
    \item \textbf{Breaking the MSD with Dynamics:} Galaxy stellar dynamics is entirely unaffected by the MSD. In principle, spatially resolved integral-field spectroscopy (IFS) kinematics can independently measure the full mass density profile of the deflector galaxy \citep[e.g.,][]{cappellari15,zhu23}. Gravitational lensing then provides a complementary and highly precise constraint on the enclosed mass and slope (if the data are good enough) modulo the MSD. Combining these two independent probes tightly constrains the total mass density profile \citep{treu_koopmans02, slacs10, shajib18, yildirim20, yildirim21}. Therefore, this joint analysis allows one to break the MSD and unambiguously constrain the total mass density profile, provided that the stellar orbital anisotropy is assumed or known to a sufficient level of precision. 
    
    \item \textbf{The Mass-Anisotropy Degeneracy (MAD):} Unfortunately, galaxy dynamics is affected by its own fundamental degeneracy. The second moment of the stellar velocities (i.e., the velocity dispersion for a non-rotating galaxy) measured from spectra only describes the component of the stellar velocity distribution along the line of sight, and the MAD states that different mass density profiles can perfectly fit the same second-order velocity moments if one simultaneously varies the anisotropy of the stellar orbits \citep{Binney1982,Gerhard1993}. The MAD cannot be constrained at all with single-aperture spectroscopy and requires spatially resolved IFS stellar kinematics and/or suitable informative priors. This is the primary strategy of the TDCOSMO collaboration outlined by \cite{shajib18} and \cite{birrer_treu21}.
    
    \item \textbf{Closing the Loop with Schwarzschild Models and Priors:} To close the loop of the lensing and dynamical degeneracies, spatially resolved kinematics and Jeans models \citep[e.g][]{cappellari08} are not sufficient on their own for nearly round, primarily pressure-supported early-type galaxies (ETGs). More general \citet{schwarzschild79} orbit-superposition dynamical models are required, which can be constrained by the full velocity histogram along the line-of-sight, rather than the second moments alone. Spatially resolved studies of nearby ETGs like the SAURON survey (\citealt{cappellari07}, \citealt[Fig.~10]{cappellari26_etgs}) show that stellar orbital anisotropy is not a universal parameter, and studies beyond the nearby Universe \citep{mozumdar26_magnus3} have extended these methods to the regime of lensing galaxies useful for cosmography. However, properly constraining Schwarzschild models requires very high-quality IFS kinematics---specifically, reliably measured, well-resolved Gauss-Hermite moments of the absorption line profile. These details cannot (yet) be obtained at intermediate redshifts ($z \sim 0.5-1.0$) and will not be obtainable until observatories like the Extremely Large Telescope (ELT) are available \citep{nguyen26_elt_sim}. This implies that we must measure the anisotropy for galaxies in the local Universe with high-quality data and use those derived anisotropies as priors for dynamical models applied to the lower-quality data at higher redshifts.
\end{enumerate}

This final step relies on the key assumption that the orbital anisotropy of massive ETGs measured in the local Universe is a valid representation of lensing galaxies up to $z \sim 1$. To give a definitive observational answer regarding the redshift evolution of anisotropy, we would fundamentally need local-quality (e.g., SAURON-like) IFS for galaxies at high redshift to robustly constrain general Schwarzschild models. Unfortunately, these kinds of data are currently out of reach, even with the James Webb Space Telescope (JWST). 

In the absence of direct high-redshift kinematic measurements, one might look to theory to estimate this evolution. However, cosmological simulations currently appear unable to reproduce in detail the anisotropy of local galaxies. The likely reason is that N-body particles in simulations are still vastly more massive than single stars, meaning artificial two-body encounters become much more frequent and artificially alter the overall orbital distribution. Thus, even though simple parametrizations of the anisotropy, e.g. constant, enable an unbiased representation of simulated galaxies, with sufficient precision to be useful for cosmography \citep{verma26, liang26_anisotropy}, empirical information is necessary for a definitive conclusion.  

In this context, the most practical way forward is to use our understanding of local data to make informative assumptions about the distant Universe. Observations of massive, round, high-velocity-dispersion ETGs---which are typical of our lens sample---indicate that they are quite close to isotropic within one effective radius ($1 R_\mathrm{eff}$) \citep[e.g.,][]{cappellari26_etgs}. It is highly unlikely that this present-day isotropy is a mere coincidence; it is much more likely driven by a stable, general physical condition, such as a state of maximum entropy. Furthermore, we know that different galaxies do not follow the exact same evolutionary path. In ongoing work analyzing 100 local galaxies with very high-quality IFS data (Fu et al., \textit{in prep.}), the homogeneity in their anisotropy is striking. It seems highly improbable that we are simply catching every single one of these galaxies at the exact same transient evolutionary phase of an overall anisotropy variation trend, where they all just happen to be isotropic. A much more natural conclusion is that these galaxies quickly converge to an isotropic state early in their formation and remain as such until the present time. Consequently, it does not matter that we observe them at different evolutionary stages (i.e., across different redshifts up to $z \sim 1$); the current local benchmarks are actually a very reliable representation of the true anisotropy of the lenses. 

We note that the situation is distinctly different for the disks of rotationally-supported ``fast rotator'' galaxies. They are definitively anisotropic (with the polar-axis velocity dispersion smaller than the radial one, $\sigma_z < \sigma_R$) \citep{cappellari07,cappellari08,Thomas2009}, a result also confirmed in the aforementioned upcoming 100-galaxy study. Crucially, however, these flattened disk systems do not suffer from the mass-anisotropy degeneracy to the same extent as nearly round galaxies. Thus, for the nearly round ETGs typical of deflectors, relying on local isotropic benchmarks provides the crucial leverage needed. Joint lensing and dynamical constraints can overcome both the MSD and MAD with the addition of informative priors on the anisotropy derived from these high-quality local observations.

TDC-25 modeled radially-symmetric spatially resolved kinematics within a spherical dynamical framework to handle the effects of the MSD in their hierarchical analysis. For time-delay lens RXJ1131-1231, KCWI kinematics from \cite{shajib23} as well as new spatially resolved kinematics from JWST-NIRSpec of  \citep{shajib26_rxj1131_nirspec} were used. Population-level constraints were introduced by radially-rebinned versions of the subset of Sloan lens ACS \citep[SLACS;][]{bolton06_slacs1, bolton08_slacs_v} lenses (henceforth SLACS/KCWI sample) presented in the first paper of this series \citep[][henceforth Paper I]{knabel24_slacskcwi}. Although spherical models are expected to be a good first-order approximation for slow rotator ETGs, this is clearly an approximation for the massive ETG population as a whole \citep[review in][]{cappellari16_review, cappellari26_etgs}. \cite{huang25_triaxiality} tested projection and triaxiality effects in Jeans anisotropic modeling \citep[JAM;][]{cappellari08, cappellari20_jeans_axis} by generating mock 3D velocity distributions from a carefully constructed synthetic lens galaxy population created from realistic instrinsic shape distributions with isothermal profiles and isotropic stellar orbits. They projected the mock kinematics through a uniform distribution of inclination angles and fitted the resulting projected line-of-sight velocity dispersions. They showed that spherical models recover the input velocity dispersions with a bias of -4\% to +1\% depending on projected axis ratio and assumption of oblateness or prolateness. They marginalized over intrinsic shape priors to derive a correction factor to the spherical approximation that can be applied as a function of projected ellipticity, which was applied to the spherical dynamical models in TDC-25 and further developed to a radially dependent correction factor for spatially resolved data. In this way, the TDC-25 dynamical models effectively corrected for and marginalized over the axisymmetric predictions from all inclination angles given the intrinsic shape priors and projected ellipticity. Following works will directly utilize axisymmetric dynamical modeling for more systems. Of primary focus to this work is the characterization of the differences in the results from directly fitted spherical and axisymmetric models with freely sampled parameters for high-quality kinematic data. Additionally, TDC-25 found an intrinsic scatter in the value of $\lambda_\mathrm{int}$ for the SLACS/KCWI sample. We will investigate this further using the axisymmetric modeling framework and the full 2D kinematic maps.

We briefly summarize the kinematic extraction of Keck KCWI IFS data in Section~\ref{sec:kinematics}. 
We describe our methods of Jeans dynamical modeling, model choices, and priors in Section~\ref{sec:dynamics}. We discuss the primary results from the models in Section~\ref{sec:results}. We summarize our conclusions in Section~\ref{sec:conclusions}. Where necessary, we assume a standard $\Lambda$CDM cosmology with $\Omega_\mathrm{m}=0.3$ and H$_0=70$ km $\mathrm{s}^{-1}$ $\mathrm{Mpc}^{-1}$. We note that measured stellar velocity dispersions are independent of cosmology.


\section{Kinematics Observations}
\label{sec:kinematics}

In Paper I, we presented spatially resolved kinematics derived from data obtained with the Keck Cosmic Webb Imager \citep[KCWI,][]{morrissey12_kcwi, morrissey18_kcwi} integral-field unit spectrograph on Keck 2 at the W.M. Keck Observatory. This sample of 14 massive ($11<\log_{\rm \rm 10} M_*<12$) lensing ETGs at redshifts z$\sim$0.15-0.35 from the SLACS survey have been uniformly strong-lens modeled by \cite{tan23_dinos}, hereafter Dinos-I. 
Kinematics were extracted from the reduced KCWI datacubes  using Penalized Pixel-Fitting method\footnote{We used v9.0.1 of the Python package from \url{https://pypi.org/project/ppxf}.} \citep[{\sc pPXF},][]{cappellari04_ppxf, cappellari17, cappellari23_ppxf}. 
Relative uncertainty in model-predicted velocity dispersions propagate directly into relative uncertainty in the MST as \citep[][Equation 20]{chen21_tdcosmo_vi}

\begin{equation}
    \mathrm{
    \frac{\delta \lambda_{int}}{\lambda_{int}} = 2 \frac{\delta \sigma}{\sigma}
    }
\end{equation}

\noindent so a 1\% bias in velocity dispersion implies a 2\% bias in $\mathrm H_0$. For this reason, we took great care to ensure our measurements are unbiased to the 1\% level and ensure they are of the precision required for cosmological inference. The kinematic maps contain data points that are spatially correlated, so we include a full covariance matrix for the uncertainty in the calculation of likelihoods. The diagonal of this matrix contains random statistical uncertainties in the bin fits as well as systematic uncertainties we estimate from the analysis we performed in Paper I. Statistical uncertainties are the formal uncertainties estimated by {\sc pPXF} and are typically on the order of a few percent for each bin. We estimated uncorrelated systematics from the spectral fitting method on the order of $1.4\%$, and we estimate off-diagonal $1\%$ covariance. See Paper I and \cite{knabel_mozumdar25_tdcxix} for a detailed budget of systematics and exploration of the spectral fitting method.


\section{Dynamical Models}
\label{sec:dynamics}

With kinematics maps in hand, we use JAM dynamical modeling software \textsc{jampy}\footnote{\url{https://pypi.org/project/jampy/}}, which solves the Jeans equations allowing for orbital anisotropy in axisymmetric or spherical alignment to dynamically describe the 3D mass profiles of the lensing deflector galaxies. See \citet{cappellari08, cappellari20_jeans_axis} for a detailed formalism in computing $\moment{v_{\rm los}^2}$ by \textsc{jampy}. JAM utilizes a Multi-Gaussian expansion \citep[MGE;][]{Emsellem94, Cappellari02} of the mass and light profiles of the galaxy using the software program \textsc{mgefit}\footnote{\url{https://pypi.org/project/mgefit/}}. \textsc{jampy} deprojects the MGE components into an oblate or prolate spheroid with an inclination angle $i$ \citep{Cappellari02}. JAM produces a model of the projected 2D stellar kinematics, which is convolved with the KCWI point-spread function (PSF, see Paper I for details) and compared with the kinematic maps from IFS data to give a log-likelihood of the data given the model: 

\begin{equation}
    \label{eq:likelihood}
    \ell(\text{data} | \text{model}) \propto \chi^2= \sum \frac{\left(\text{model} - \text{data}\right)^2}{\text{error}^2}
\end{equation} 

This log-likelihood is minimized using Markov Chain Monte Carlo methods (MCMC).

\subsection{Mass profile}

Power law elliptical mass density (PEMD) models have been shown to give a good description of the total (dark and baryonic) mass density profiles of ETGs through joint lensing and dynamical \citep{Treu2004} and purely dynamical studies \citep{cappellari15}. PEMD models in lens model formalism are described by the logarithmic radial slope of the mass profile $\gamma$, the Einstein radius $\theta_{E}$ that anchors the normalization of the profile, and the projected axis ratio $q$, where the projected surface mass density is proportional to the convergence profile $\kappa_\mathrm{PL} ({\theta})$:

\begin{equation}
    \mathrm{
    \Sigma_\mathrm{PL} ({\theta}) \propto \bigkappa_{ PL}({\theta}) = \frac{3 - \gamma}{2} \left( \frac{\theta_{E}}{\theta} \right ) ^ {\gamma - 1}
    }
    \label{eq:power_law}
\end{equation}

\noindent where $\theta$ is the projected angular radius $\theta=\sqrt{q\theta_1^2+\theta_2^2/q}$ at coordinates ($\theta_1$,$\theta_2$). $\gamma=2$ for an isothermal profile. $q=1$ for spherical models. The Einstein radius $\theta_\mathrm{E}$ is defined at the intermediate axis, so that the major axis of the PEMD ellipse would be $\theta_\mathrm{E} / \sqrt{q}$. For axisymmetric models, we fix the observed axis ratio of the mass to that of the observed surface brightness profile of the lens galaxy. We also include a central supermassive black hole as a point mass at the center of the galaxy. This form is common to all models in this work and is the base profile that undergoes the MST.

In practice, from the sampled input parameters, $\bigkappa_\mathrm{PL}$ is first constructed as a circular pure power law as in Equation \ref{eq:power_law}. This profile undergoes the total (internal and external) MST through

\begin{equation}
    \mathrm{
    \bigkappa_{MST,\: tot}({\theta}) = \lambda_{MST,\: tot} \: \bigkappa_{PL}({\theta}) + (1-\lambda_{MST,\: tot})
    }
    \label{eq:mst_total}
\end{equation}

\noindent The fully transformed profile $\bigkappa_\mathrm{MST,\: tot}$ is then converted to surface mass density and fit with MGE. Through these steps, the mass density is described as a 1D circularized profile that will be converted to its true elliptical profile after being converted with MGE. 

We approximate $\bigkappa_\mathrm{MST,\: tot}$ by two subsequent transformations (first the internal and then the external) in the following equations. We use Equation \ref{eq:mst_int} to transform the pure power-law convergence profile from Equation \ref{eq:power_law} by the internal MST, to get the convergence profile $\bigkappa_\mathrm{MST,\: int}$, distinct from the fully transformed version $\bigkappa_\mathrm{MST,\: tot}$.  

\begin{equation}
    \mathrm{
    {\bigkappa_{ MST,\: int}}({\theta}) = \lambda_{int} \: \bigkappa_{ PL}({\theta}) + (1-\lambda_{int}) \: \bigkappa_s(\theta)
    }
    \label{eq:mst_int}
\end{equation}

\begin{equation} 
    \mathrm{
    \bigkappa_s(\theta) = \frac{{\theta_s}^2} {\theta^2 + {\theta_s}^2 }
    }
    \label{eq:mst_trunc}
\end{equation}

\noindent The truncation function $\bigkappa_\mathrm{s}$ (Equation \ref{eq:mst_trunc}) is necessary to approximate a real galaxy with finite total mass when considering only the internal MST. $\rm \theta_s$ is not physical parameter, but it must be carefully chosen to closely approximate the pure mass sheet in the inner regions of the galaxy such that the lensing images are insensitive to the effect of the changes in the mass profile. We discuss the selection of values for this parameter in Section \ref{sect:priors_bounds}.

We transform $\rm \bigkappa_{MST,\: int}$ by the external MST, defined by $\rm \lambda_{ext}=1-\bigkappa_{ext}$, where $\rm \bigkappa_{ext}$ is the median external convergence calculated for the lens from line-of-sight studies \citep[e.g.,][]{wells24_tdcxv}. This transformation does not require the truncation function (Equation \ref{eq:mst_trunc}) because $\rm \kappa_{ext}$ is a line-of-sight effect that does not affect the gravitational potential in which the stellar population of the deflector galaxy orbits. The result $\rm \bigkappa_{MST,\: tot}^{eff}$ is an effective approximation to the fully transformed non-truncated convergence profile $\rm \bigkappa_{MST,\: tot}$ (Equation \ref{eq:mst_total}). 

\begin{equation}
    \mathrm{
    \bigkappa_{MST,\: tot}^{eff}\left(\theta\right) = \lambda_{ext} \: {\bigkappa_{ MST,\: int}\left(\theta\right)} + \left(1 - \lambda_{ext}\right)
    }
\end{equation}

This profile is then multiplied by the cosmology-dependent critical surface mass density $\Sigma_\mathrm{crit}$ to give physical units. 

\begin{equation}
    \mathrm{
    \Sigma_{MST,\: tot}^{eff}\left(\theta\right) = \bigkappa_{MST,\: tot}^{eff}\left(\theta\right) \: \Sigma_{crit}
    }
    \label{eq:surface_mass_dens}
\end{equation}

\begin{equation}
    \Sigma_\mathrm{crit} = \frac{c^2 D_\mathrm{s}}{4 \pi G D_\mathrm{ds} D_\mathrm{d}}
\end{equation}

\noindent where $D_\mathrm{d}$, $D_\mathrm{s}$, and $D_\mathrm{ds}$ are the angular diameter distances to the main foreground deflector, to the background source, and between the foreground deflector and the background source. 

The central black hole point mass will contribute to the mass enclosed within the Einstein radius, so we must scale the surface mass density profile to compensate. We integrate the surface mass density within the Einstein radius to get the total mass enclosed (excluding the black hole).

\begin{equation}
    M_\mathrm{enc} = 2 \pi \int_{0}^{R_\mathrm{E}} \Sigma_\mathrm{MST,\: tot}^\mathrm{eff}\left(R \right) \: R \,dR 
\end{equation}

\noindent where $R_\mathrm{E} = D_\mathrm{d} \: \theta_\mathrm{E}$ is the physical perpendicular radius of the Einstein radius and $R = D_\mathrm{d} \: \theta$ is the physical perpendicular radius at which the function is evaluated. In order to conserve the enclosed mass within the Einstein radius when we add the point mass central black hole, we need to rescale the surface mass density profile.

\begin{equation}
    \Sigma_\mathrm{MST,\: tot, \: BH}^\mathrm{eff} = \frac{M_\mathrm{enc} + M_\mathrm{BH}}{M_\mathrm{enc}} \: \Sigma_\mathrm{MST,\: tot}^\mathrm{eff}
\end{equation}

Finally, the radial profile of the circularized surface mass density profile $\rm \Sigma_{MST,\: tot, \: BH}^{eff}$ is sampled at 1000 logarithmically spaced radii from 0.01 arcsec to 20 $R_\mathrm{eff}$ and fit with \textsc{mge\_fit\_1d} using up to $N=30$ total Gaussian components, with the $j$-th component defined by its total counts (integral of the 1D Gaussian) $M_j$ and dispersion $\sigma_j$:

\begin{equation}
    \Sigma_\mathrm{MST,\: tot, \: BH}^\mathrm{eff}\left(\theta\right) \approx \Sigma_\mathrm{MGE}\left(\theta\right) = \sum_{j=1}^{N} \frac{M_j}{\sqrt{2 \pi} \sigma_j} \exp{\left[\frac{1}{2}\frac{\theta^2}{\sigma_j^2}\right]}
    \label{eq:mge_approx}
\end{equation}

\noindent We ensure the MGE profile closely approximates the analytical mass density profile by setting a threshold of 5\% for the allowed residual for each sampled radius within a range of sensitivity of the kinematics. Residuals in the MGE fit generally occur at the smallest and largest radii sampled. The dynamical fit to the kinematic map is not sensitive at radii that are much smaller than the KCWI spaxel size or much larger than the outermost spaxel. For the lower end of the range for our residual test, we take the larger value of 0.01 arcseconds or the minimum $\sigma_{j}$) from the first fit, which is always much smaller than the spaxel size of 0.1457 arcseconds. The upper end of the range is 30 arcseconds. If this step results in residuals greater than 5\%, we double the number of Gaussian components to give it more flexibility. \textsc{mge\_fit\_1d} automatically optimizes the number of Gaussians and never needs the maximum number for our fits. If the residual is still greater than 5\% at any sampled radius in the range, we double the number of sampled points and the maximum sampled radius, which serves to move the higher-residual extreme ends of the fit further outside the range we care about. If this threshold still is not satisfied, this reflects a model that is not easily defined by MGE and is therefore rejected. If the threshold is satisfied, the profile is converted to an elliptical surface mass density profile by dividing the Gaussian dispersions $\sigma_{j}$ by the square root of the projected axis ratio $q$ (which we apply uniformly to all MGE components in the fit, thus fixing $q$ at all radii). The JAM models take as inputs the peak surface mass density $\Sigma_{j,\mathrm{0}}$, the dispersion along the major axis $\sigma_{j,\mathrm{maj}}$, and the axis ratio $q$.

\begin{equation}
    \sigma_{j,\mathrm{maj}} = \frac{\sigma_{j}}{\sqrt{q}}
    \label{eq:mge_sigma_maj}
\end{equation}

\begin{equation}
    \Sigma_{j,\mathrm{0}} = \frac{M_j}{\sqrt{2 \pi} \sigma_j}
    \label{eq:mge_peak}
\end{equation}

\noindent These parameters define the surface mass density that JAM will deproject through an inclination angle that is defined by the intrinsic shape of the galaxy. 

The stellar dynamical tracer population is deprojected through the same inclination angle to give the distribution of stars whose kinematics are dictated by the gravitational potential of the intrinsic dynamical mass model and the anisotropy. The stellar light is translated to MGE components from the lens model's deflector S\'{e}rsic or double S\'{e}rsic light profile. Dinos-I deblended the lens deflector light from contaminating flux from the background source arcs and accounted for the PSF to recover the unblurred light profile. The observed axis ratio $q$ of the deblended deflector light profile sets the axis ratio uniformly for all MGE components for the light and mass. Therefore, the light and mass are fixed to the same center and axis ratio. Otherwise, the profiles are distinct.

Now we have a 3D mass profile and 3D tracer population. JAM solves the Jeans equations with orbital anisotropy and integrates along the line of sight for each spaxel location ($x_i,y_i$ for spaxel $i$) in the kinematic maps to produce line-of-sight moments of velocity on the spaxel grid, which is convolved with the KCWI PSF \citep[see Appendix A of][]{cappellari08}. The second moment of velocity $\moment{v_{\rm los}^2}$ is closely comparable to the measured $V_\mathrm{rms}$. In order to compare with the spatially binned data $V_{\mathrm{rms},j}$ for bin $j$, we calculate the corresponding model line-of-sight velocity second moment $\moment{v_{\rm los}^2}_j$ 
\begin{equation}\label{eq:v_los}
	\moment{v_{\rm los}^2}_j = \frac{1}{F_j} \sum_i F_{i,j} \moment{v_{\rm los}^2}_{i,j} .
\end{equation}
\noindent where $\moment{v_{\rm los}^2}_{i,j}$ is the line-of-sight second velocity moment at spaxel $i$, and the average is weighted by the model flux weights $F_{i,j}$ for each spaxel $i$ belonging to bin $j$. $F_j = \sum_i F_{i,j}$ is the total of bin $j$. All terms in Equation~\ref{eq:v_los} are PSF-convolved.

The residual $V_{\mathrm{rms},j} - \sqrt{\moment{v_{\rm los}^2}_j}$ is normalized by the bin uncertainty, and the sum of the square normalized residuals is proportional to the log-likelihood as in Equation \ref{eq:likelihood}. We add the log of the prior probability, and this sum is maximized by the sampling algorithm in a Bayesian framework. We use MCMC with python package {\sc Emcee}\footnote{\url{https://pypi.org/project/emcee/}} \citep{foreman13_emcee} to sample the parameters and estimate their posterior probability distribution.

\subsection{Model choices}

Our baseline model is a joint lensing and dynamical model that assumes spherical symmetry in the mass profile and the stellar kinematic tracer distribution for comparison with previous joint lensing and dynamical studies (e.g., aperture spectroscopy and radially binned IFS kinematics; especially as in TDC-25). Spherical models have been shown to be appropriate for the most massive nearby ETGs, which are classified as ``slow rotators" \citep{emsellem07, emsellem11} and are typically approximately round and only weakly triaxial, with ellipticities $\epsilon = 1 - q<0.4$ \citep[see review in][]{cappellari26_etgs}. We showed in Paper I that most (11/14) of our SLACS/KCWI sample is consistent with being slow rotators in kinematic properties and observed ellipticity through ($V/\sigma,\epsilon$) and ($\lambda_{\rm R},\epsilon$), which are diagnostics that describe the relative significance of rotational versus dispersion velocities. The other three objects have kinematic properties near the overlap boundary between fast and slow rotators. Although these galaxies exhibit clear rotation, their high velocity dispersions imply they have massive bulges that dominate the lensing behavior. These objects will be particularly valuable for determining any biases arising from the baseline spherical Jeans model. For these spherically symmetric models, we circularize the MGE decomposition of the surface brightness profile and assume a spherical power-law mass model ($q=1$ in Equation \ref{eq:mge_sigma_maj}).

While spherical models are appropriate as a first-order approximation, real galaxies are better described with a symmetry axis. Axisymmetric models more accurately deproject the observed surface brightness profile to an intrinsic 3D shape of the tracer distribution, utilizing the Multi-Gaussian Expansion (MGE) formalism and knowledge of the intrinsic shapes of nearby ETGs.
Within these axisymmetric models, we test anisotropy prescriptions in which stellar velocity ellipsoids are aligned either with spherical coordinates (radially toward the galaxy center) or with the cylindrical symmetry axis. The orientation of the velocity ellipsoid is set by the gravitational potential and is therefore not expected to depend on galaxy assembly history. Orbit integrations indicate that galaxies are generally expected to have velocity ellipsoids aligned in spherical coordinates \citep[sec.~2.1]{cappellari26_spectral_jam}. 

In contrast, the anisotropy itself---namely the flattening of the velocity ellipsoid and its spatial variations---can depend on galaxy evolution. In particular, some evidence suggests that the angular variation of the ellipsoid flattening depends on kinematic morphology \citep{cappellari07}. For slow rotators, the elongation of the velocity ellipsoid is expected to remain relatively constant with polar angle. Fast-rotating galaxies with stellar disks show more complex behavior: although their velocity ellipsoid is approximately spherically-aligned throughout the galaxy, its shape changes with angular position. As discussed in \citet[sec.~5.5]{cappellari26_spectral_jam}, the major axis is expected to be radially aligned in the equatorial plane but to become orthogonal to the radial direction (tangential) near the symmetry axis.
This behavior in fast rotators leads to a global anisotropy that, to first order, resembles a simple flattening of the velocity ellipsoid in the $z$-direction. For this reason, assuming anisotropy in cylindrical coordinates is generally an appropriate approximation for the kinematics of fast rotators, while spherical alignment remains preferred for slow rotators. In practice, however, dynamical models of the ATLAS$^{\rm 3D}$ \citep[sec.~8.6]{cappellari20_jeans_axis} and MaNGA DynPop \citep{zhu23} samples show that both alignments produce highly consistent mass model properties (within a few percent) and are statistically indistinguishable in terms of $\chi^2$. Notably, the slope of the mass profile is insensitive to the assumed alignment of the velocity ellipsoid for axisymmetric models.

The orbital anisotropy of ETGs is expected to vary radially out to large radii. However, studies of nearby galaxies have shown them to be well-characterized by ``constant" or spatially-uniform anisotropy, i.e., $\beta_\mathrm{ani}\left(r,\theta\right) = \beta_\mathrm{ani}$ for all values of $r$ and $\theta$. Using \citet{schwarzschild79} orbit modeling methods that allow for radial variation of $\beta_\mathrm{ani}$ and IFS data of nearby galaxies with exceptionally high resolution, \cite{cappellari26_review} shows flat profiles close to isotropic in the range $R_\mathrm{eff} / 30 < r < R_\mathrm{eff}$, where our data is sensitive. \cite{verma26} studied anisotropy profiles using simulated datasets and found that constant anisotropy does not significantly bias the results compared to radially varying models. We therefore assume constant anisotropy in all our models and leave tests of radially varying anisotropy models to future work.

For axisymmetric models, we fix the intrinsic axis ratio of the mass model to that of the stellar kinematic tracer distribution, and we fit both simultaneously. Because most ETGs are oblate and axisymmetric (or only weakly triaxial), the standard approach is to assume oblate surface mass and brightness profiles in deprojection. Cases of truly prolate rotation are nonexistent in fast rotator ETGs \citep{Krajnovic2011} and rarely observed even in the slow rotator population \citep{li2018_manga}, except for the most extreme masses \citep{Krajnovic2018}; galaxies elongated in a more prolate manner tend to be more triaxial than purely prolate. 

For systems exhibiting clear rotational velocities, we estimated the kinematic major axis using the procedure {\sc fit\_kinematic\_pa}, which is part of the \textsc{PaFit} package\footnote{\url{https://pypi.org/project/pafit/}} \citep[app.~C]{krajnovic_06_fitkinpa}. By comparing the position angle of the rotation axis to the minor axis of the surface brightness contours, we can gauge deviations from axisymmetry. All objects in our sample exhibit misalignments of $\Delta \rm PA < 45^{\circ}$, which is consistent with an oblate shape\footnote{In Paper I, object J1250+0523 was reported to have an almost $90^{\circ}$ misalignment; this has since been remeasured using the correct photometric PA and is now consistent with the oblate case.}. While nontrivial misalignments are suggestive of a slight triaxial nature that axisymmetric JAM cannot fully model, explicit treatment of deviations from axisymmetry is beyond the scope of this paper.

The dynamical models are prepared with the spatially binned kinematic map $V_\mathrm{rms}$ constructed using pPXF from Paper I and MGE decompositions of the deblended surface brightness profile of the deflector light from the lens model, aligned along the photometric major axis. We specify mass model, 
symmetry assumption, alignment of the velocity ellipsoid, redshifts of the foreground deflector and lensed background source, the measured KCWI PSF FWHM (see Paper I), 
error estimates and covariance terms, bounds, priors, and the sampler arguments for MCMC (number of walkers, steps, and dimensions).
We additionally test all configurations with purely dynamical constraints, i.e., without including lens model information as priors on the fitted parameters. All our baseline models explicitly include the MST parameter, i.e. $\lambda_\mathrm{int}$ is not fixed to 1. We performed tests with $\lambda_\mathrm{int}=1$ for the sake of comparison; unless otherwised stated, all models we discuss have freely fitted $\lambda_\mathrm{int}$.

In order to quantify model preference, we compare Bayesian information criteria (BIC) as defined in \citep{schwarz78_bic}: $\mathrm{BIC} = k \: \ln(N) - 2 \ln\left(\hat{L}\right)$, where $k$ is the number of free fit parameters, $N$ is the number of fitted data points, and $-2 \ln\left(\hat{L}\right) = \chi^2_\mathrm{best}$ is the log-likelihood or $\chi^2$ of the best fitting parameters as in Equation \ref{eq:likelihood}. We calculate the error on BIC $\rm \sigma_{BIC}$ from the sample $\chi^2$ distribution. Given a dataset, a model with a lower BIC is considered to be a better fit than another model of the same data with a higher BIC. For models 1 and 2,  if $\Delta \rm BIC_{12} \equiv BIC_1 - BIC_2 > \sigma_{\Delta BIC, 12}$, where $\rm BIC_1 > BIC_2$ and $\rm \sigma_{\Delta BIC, 12} = \sqrt{\sigma_{BIC,1}^2 + \sigma_{BIC,2}^2}$, this indicates a distinguishable preference for model 2 over model 1.

\subsection{Fit parameters, priors, and bounds}\label{sect:priors_bounds}

Parameter bounds and default priors are given in Table \ref{tab:models}. The number of fitted parameters depends on the assumption of symmetry. All models share the following parameters: 1) $\gamma$ - logarithmic slope of pure power-law mass profile, 2) $\theta_\mathrm{E}$ - Einstein radius for normalization of pure power-law mass profile, 3) $\log M_\mathrm{BH}$ - central supermassive black hole mass, and 4) $\rm \sigma_{tan} / \sigma_{rad}$ 
the constant anisotropy ratio $\sqrt{1 - \beta_\mathrm{ani}}$. The slope $\gamma$ is significantly more constrained by the dynamical model than by the lens model. This is expected because the SLACS imaging data were designed to confirm lensing and measure the Einstein radius, not to infer the slope to high precision, contrary to the deeper exposures obtained for the time-delay lenses \citep[see also discussion in][]{tdcosmo25_milestone}. Early tests showed that the inclusion of the lens model posterior on $\gamma$ was uninformative to the dynamical model, so we use a uniform prior in the bounds given in Table~\ref{tab:models}. The Einstein radius $\theta_\mathrm{E}$ is well constrained by lensing, so we use Gaussian priors from the posteriors of the lens models, including both the statistical and systematic uncertainty estimates added in quadrature. For dynamics-only models, the Einstein radius is fitted with uniform priors. The central black hole mass $\log M_\mathrm{BH}$ is fitted with a Gaussian prior with a mean estimated using the $M_\mathrm{BH} - \sigma_\mathrm{eff}$ scaling relation from~\citet[][equation 7 therein]{kormendy_ho13}, and we assume intrinsic scatter of 0.3 dex in $M_\mathrm{BH}$.

The anisotropy ratio is defined as $\sigma_\mathrm{tan}/\sigma_\mathrm{rad} = \sigma_{z}/\sigma_{R}$ for a cylindrically-aligned ellipsoid and $\sigma_\mathrm{tan}/\sigma_\mathrm{rad}=\sigma_{\theta}/\sigma_{r}$ for a spherically-aligned velocity ellipsoid. For the spherical and spherically-aligned axisymmetric models, we take a Gaussian prior of $1.0\pm0.07$ from the mean and standard deviation of the sample of lenses modeled with Schwarzchild models in \citep[][]{cappellari26_etgs}. \cite{mozumdar26_magnus3} used this prior and compared with results using a flat prior, finding that while the choice of prior influences the specific best-fit anisotropy and mass parameters, it has a negligible impact on the derived total density slope of the mass profiles of the galaxies. Cylindrically-aligned models have an empirically determined upper bound at 1 that does not allow tangential anisotropy in the $z$ direction in order to avoid a degeneracy with the inclination angle, and spherically-aligned models are bound at 2 \citep[as in][]{zhu23}. Spherical and spherically-aligned models are bound between 0 and 2 since they are not subject to any inclination effect.

Axisymmetric models also include the intrinsic axis ratio $q_\mathrm{intr}$ that describes both the mass and light profiles. We fit this parameter as opposed to the inclination because it allows us to leverage prior information from the intrinsic shapes of nearby ETGs (e.g., ATLAS$^{3\rm D}$, Manga DynPop). In Fig.~4 of Paper I, we presented the kinematic maps with contours showing the MGE decompositions of the B-spline models to the foreground deflector light profile. These models allowed each Gaussian component to have its own axis ratio. In nearby galaxy studies radial gradients in the axis ratio are commonly found. However, the spatial resolution of our sample is better suited to a fixed axis ratio to avoid overfitting. In Dinos-I, the surface brightness profiles were fitted with at most two S\'ersic profiles, each with its own axis ratio. For the observed surface brightness profiles in our JAM models, we translate the S\'ersic profiles with MGE and use the axis ratio of the dominant light profile as the common axis ratio $\rm q_{obs}$ for all Gaussian components. For spherical models, we circularize the observed surface brightness profile to the intermediate axis by multiplying the Gaussian widths $\sigma_{j,\mathrm{maj}}$ by $\sqrt{q_\mathrm{obs}}$ as in Equation \ref{eq:mge_sigma_maj}. 
Each of the MGE components of the observed surface brightness profile is deprojected through an inclination angle to the intrinsic shape in 3D space described by $q_\mathrm{intr}$. We take a Gaussian prior on $q_\mathrm{intr}$ from nearby slow rotators of $0.74\pm0.08$ from \cite{li2018_manga} with bounds between 0.051 and $q_\mathrm{obs}$ for the object. This prior helps break the degeneracy between the inclination and the anisotropy profile under assumptions of axisymmetry. The inclination angle $i$ is calculated from $q_\mathrm{obs}$ and ${q_\mathrm{intr}}$ as

\begin{equation}
    \sin{i} = \sqrt{ \frac{ 1 - {q_\mathrm{obs}}^2 } { 1 - {q_\mathrm{intr}}^2 } }
    \label{eq:inc}
\end{equation}

\noindent We also tested a prior with a flatter mean ${q_\mathrm{intr}}$ of $0.25\pm0.14$ from nearby fast rotators from \cite{weijmans2014_qintr}. For the three objects with strong rotation that we identified to be on the boundary between slow and fast rotators, the fits with this prior were significantly worse than with the slow rotator prior. For our final models, we use the slow rotator prior for all objects.

Our main models, which include the MST, have two additional parameters: 1) $f_\mathrm{MST}$ - a parameter in range [0,1] that is a function of the internal MST parameter $\rm \lambda_{int}$, and 2) $a_\mathrm{MST}=\theta_\mathrm{s} / \theta_\mathrm{E}$ - truncation radius normalized by the Einstein radius that forces the internal mass sheet to be 0 at large distances from the galaxy (as in Equation \ref{eq:mst_trunc}). $f_\mathrm{MST}$ maps to $\lambda_\mathrm{int}$ as

\begin{equation}\label{eq:k_par}
    \lambda_\mathrm{int} = \lambda_\mathrm{lo} + f_\mathrm{MST} (\lambda_\mathrm{hi} - \lambda_\mathrm{lo}) 
\end{equation}

\noindent where the lower bound $\rm \lambda_{lo}$ is a conservative lower limit on $\rm \lambda_{int}$ based on \cite{birrer20_tdcosmo_iv} and the upper bound $\rm \lambda_{hi}$ is dependent upon the sampled model realization. $\rm \lambda_{hi}$ is the minimum of either 1.2 or the maximum allowed value for $\rm \lambda_{int}$ for the given power-law profile such that the transformed convergence profile from Equation \ref{eq:mst_int} is not negative at any radius. This maximum is sensitive to the choice of the MST truncation (or core) radius $\rm \theta_s$, where for a given value of $\rm \theta_s$, a steeper profile with less total mass will give negative convergence for fixed values of $\rm \lambda_{int}>1$. This is an unphysical constraint on the value of $\rm \lambda_{int}$ that is wholly dependent on the arbitrary choice of the truncation radius that is necessary within the analytic form of the approximate mass sheet. For this reason, we fit $a_\mathrm{MST} = \theta_\mathrm{s} / \theta_\mathrm{E}$ as a nuisance parameter that will allow flexibility of a given power-law realization to include $\rm \lambda_{int} > 1$ without resulting in negative convergence. The bounds on $a_\mathrm{MST}$ of [5, 10] are motivated by \cite{birrer20_tdcosmo_iv}, which showed that the core radius $\rm \theta_s > 5 \theta_E$ is typically sufficiently far away from the lensed arcs to maintain the MST's inability to be detected by the lens model. This avoids unphysical models and allows these parameters to be explored uniformly within the allowed parameter space given the other sampled parameters. For any truncation radius, a given transformed convergence profile is well-fit by the MGEs within $\sim10\: R_\mathrm{eff}$, and the effect on the resulting kinematic model is well below 1\%.

The external convergence ($1 - \lambda_\mathrm{ext}$) is fixed at the median value of the posterior from line-of-sight studies as in TDC-25. We treat it as a piece-wise Gaussian about the median of the true posterior, where $\sigma$ on either side of the median is given by the 16th and 84th percentiles. The uncertainty of $\kappa_\mathrm{ext}$ is folded into the Einstein radius prior because $\theta_\mathrm{E}$ is purely degenerate with the scaling effect of the external mass sheet.
The uncertainties from the lens model on $\theta_\mathrm{E}$ are very small (less than 1\%). This is expected, since lensing very tightly constrains projected mass enclosed within the measured $\theta_\mathrm{E}$, including the galaxy and external mass along the line of sight. This is not the same thing as uncertainty on the absolute scaling of the galaxy's mass profile, which is susceptible to the MSD.
The scaling effect of $\kappa_\mathrm{ext}$ on the galaxy mass normalization after undergoing the MST is

\begin{equation}
    \mathrm{
    \theta_{E,MST} \propto \left ( 1 - \kappa_{ext} \right ) ^ \frac{1}{ 1 - \gamma } 
    }
\end{equation} 

\noindent so

\begin{equation}
\begin{split}
\frac{\delta\theta_\mathrm{E,MST}}{\theta_\mathrm{E,MST}} = \Biggl( & \left[ \frac{1}{\gamma - 1} \frac{\delta\kappa_\mathrm{ext}}{1 - \kappa_\mathrm{ext}} \right]^2 + \left [ \frac{\delta\theta_\mathrm{E}}{\theta_\mathrm{E}} \right]^2 \\ & + \left [ \ln (1 - \kappa_\mathrm{ext}) \frac{\delta\gamma}{(\gamma - 1)^2} \right]^2  \Biggr)^{1/2}
\end{split}
\end{equation}

\noindent $\gamma$ is close to 2 (isothermal), and $\kappa_\mathrm{ext}$ is close to 0 for all objects. All three terms will be quite small for all objects, and the second term is negligible for almost all cases. Added in quadrature as above, the uncertainty is inflated to a level of 1-3\% in the prior we use on the dynamically fitted (mass-sheet susceptible) Einstein radius.

\begin{table*}[]
    \centering
    \begin{tabular}{l l|l|l}
    parameter &   &   bounds   &    mean, $\sigma$\\
    \hline

    \textbf{common parameters} & & &  \\

    slope & $\gamma$  & [1.6, 2.6]   &  uniform   \\
    
    Einstein radius & $\theta_\mathrm{E}$ [arcsec] & [0.7, 2.0]   &  lens posterior    \\

    black hole mass & $\log M_\mathrm{BH}$ [$M_{\odot}$] & [7.5, 10] & 0.3 \\

    \hline

    \textbf{spherical tracer} & & & \\
    
    anisotropy & $\sigma_{\theta}/\sigma_r$ & [0.0 2.0] & 1.0, 0.07   \\ 

    \hline

    \textbf{axisymmetric tracer} & & \\
    
    intrinsic shape & $q_\mathrm{{intr}}$ & [0.051, $\mathrm{q_{obs}}$]  &  0.74, 0.08     \\

    anisotropy & sph $\sigma_{\theta}/\sigma_r$ & [0.01, 2.0] & 1.0, 0.07 \\ 

    & cyl $\sigma_{Z}/\sigma_R$ &  [0.01, 1.0] & uniform \\

    \hline

    \textbf{mass sheet} & & \\

    internal mass sheet & $\mathrm{ \lambda_{int} }$ & [0.8, 1.2] & uniform$\dagger$ \\

    MST scale radius & $\mathrm{ a_{MST}}$ & [5, 10] & uniform \\

    \end{tabular}
    \caption{Priors and bounds for JAM models. $\dagger\lambda_\mathrm{int}$ is uniformly sampled in the transformed parameter space defined by $f_\mathrm{MST}$ and is dependent on the sampled model. See Equation \ref{eq:k_par} and Section \ref{sect:priors_bounds}.}
    \label{tab:models}
\end{table*}


\section{Results}
\label{sec:results}

Model and data comparisons, as well as plots for each object in the sample showing the primary parameters of concern to our study are given in Appendix \ref{sec:corner_plots}.

\subsection{Constraints on MAD and MSD}\label{sec:mad_msd_constraints}

All models achieve constraints on the anisotropy and mass profiles. Table \ref{tab:res_errors} shows the average errors per object for $\beta_\mathrm{ani}$, $\gamma$, $\theta_\mathrm{E}$, and $\lambda_\mathrm{int}$, marginalized over the models with BIC weights. The dynamics-only models are remarkably constraining on their own, even on the shape of the mass profile when including the MST parameter $\lambda_\mathrm{int}$ as a free parameter. Uncertainties in $\theta_\mathrm{E}$ and $\lambda_\mathrm{int}$ are significantly improved in the joint models and are insensitive to the anisotropy prior; if the anisotropy prior is removed, the mass normalization is still tightly constrained by the lensing enclosed mass and the kinematics. Average precision on the MST parameter $\rm \lambda_{int}$ for dynamics-only models is 8.2\%, improved to 5.8\% for the joint models, confirming the success of the methodology detailed in Section \ref{sec:dynamics}. There is no improvement on the uncertainty in $\beta_\mathrm{ani}$ or $\gamma$ from the joint models, since the constraints on the anisotropy come primarily from the prior, and the slope is more precisely measured from the dynamics than the lensing. As discussed in Section \ref{sec:dynamics}, even with the joint constraints, informative priors on the anistropy from local galaxy studies are needed to break the residual MAD. However, the MSD and MAD are both overcome by the combination of lensing, dynamics, and the informative prior.

\begin{figure*}
    \centering
    \includegraphics[width=\linewidth]{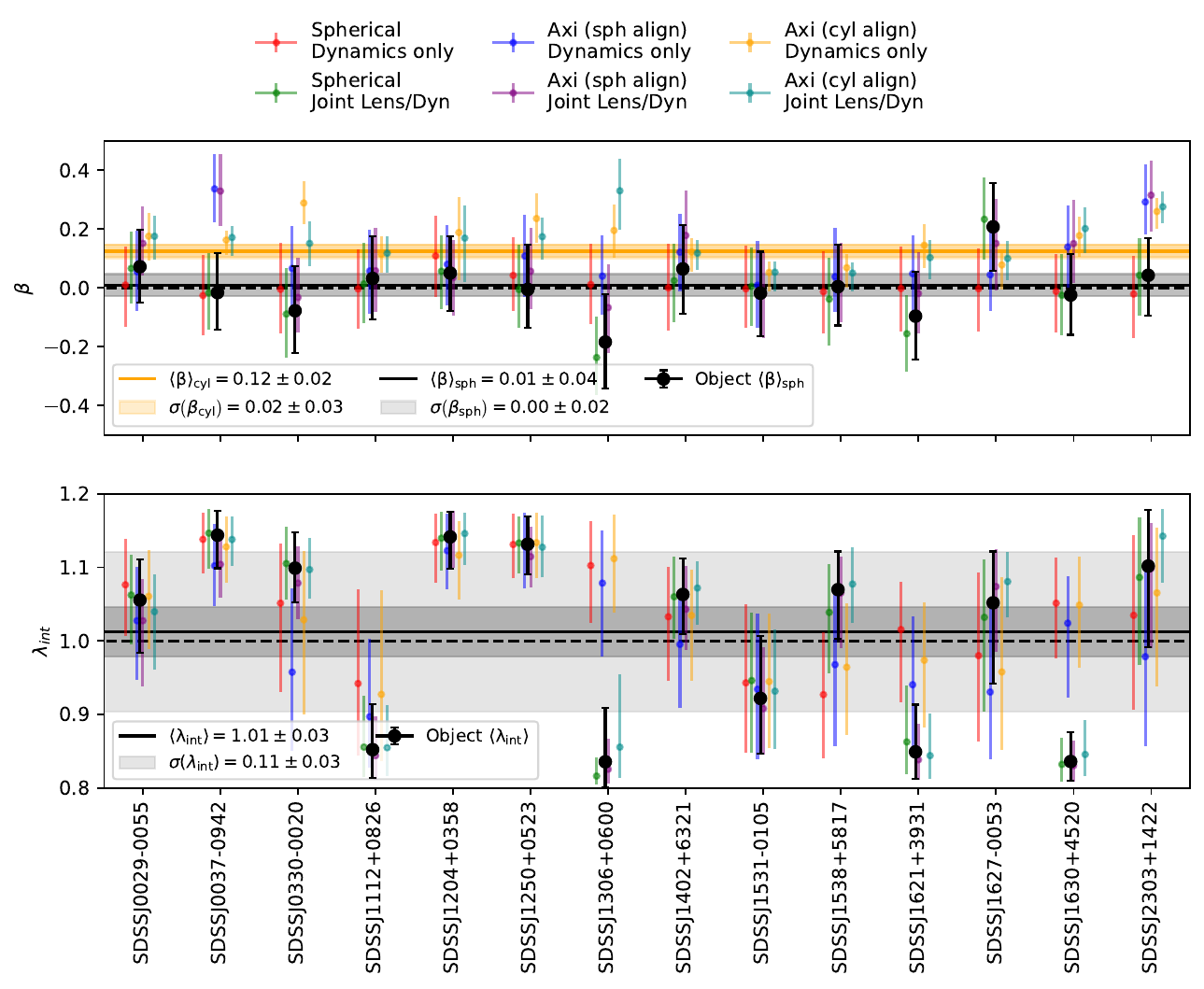}
    \caption{All models (dynamics-only and joint) are shown, but only the joint models are included in the average. Parameters are averaged with BIC weights for each object. Sample means are shown as solid horizontal lines, error on the sample mean is the filled dark region, and intrinsic scatter of the sample is the lighter filled region. \textit{Upper panel}: anisotropy parameter $\beta_\mathrm{ani}$. Black circles show the mean value for joint spherical and spherically-aligned axisymmetric models for each object, and the black line and fill region are the sample mean and error. The orange line and fill region show the sample and error on the mean for the cylindrically-aligned axisymmetric models. The intrinsic scatter happens to be identical to the error on the mean, and the shaded regions overlap. \textit{Lower panel}: MST parameter $\rm \lambda_{int}$. Black circles show the mean value for all three joint models for each object. Black horizontal line and fill regions show the mean, error on the mean, and scatter over the sample.}
    \label{fig:beta_lambda_plot}
\end{figure*}

We show a summary of the results for the anisotropy and MST parameters in Figure~\ref{fig:beta_lambda_plot}. For each object on the x-axis, we plot the value and uncertainty for each of the six models. Red, blue, and yellow correspond to dynamics-only models, while green, purple, and cyan correspond to joint lensing/dynamical models. The weighted mean of the joint lensing/dynamical models (excluding the dynamics-only models) for each object are calculated using the BIC formalism described in Section \ref{sec:bic_model_preference}, and are plotted as black circles. Error bars on the BIC-weighted mean include formal and systematic errors estimated with BIC weighting. Sample means, errors on the mean, and scatter over the sample are shown as horizontal solid lines and filled regions.

The anisotropy parameter $\beta_\mathrm{ani}$ is shown in the upper panel. For the sample mean, error, and scatter, shown as horizontal lines and filled regions, we separate the spherically-aligned models (pure spherical and axisymmetric spherical) from the cylindrically-aligned models. The mean of the spherically-aligned models for each object is shown as a black circle. The spherically-aligned models have a mean of $\rm \left \langle \beta \right \rangle_{sph} = 0.01\pm0.04$ with no intrinsic scatter, and are consistent with being drawn from the anisotropy prior we used for those models (See Table \ref{tab:models} and Section \ref{sect:priors_bounds}) with a Kolmogorv-Smirnov statistic of 0.24 and p-value of 0.35. The cylindrically-aligned models are on average slightly radial: $\rm \left \langle \beta \right \rangle_{cyl} = 0.12\pm0.02$ with a scatter of $0.02\pm0.03$. This is expected since our prior requires $\beta^\mathrm{cyl}_\mathrm{ani} > 0$. 

The bottom panel of Figure \ref{fig:beta_lambda_plot} shows the sample level summary of the MST parameter $\rm \lambda_{int}$. While the mean of the sample is $\rm \left \langle \lambda_{int} \right \rangle = 1.01\pm0.03$, and thus consistent with no MST within the errors, the scatter of $0.11\pm0.03$ indicates the need for deviations from the pure power law for individual lenses. Other joint lensing/dynamical studies \citep[e.g.,][]{etherington23} have noted evidence for deviations from the pure power law form. Dinos-I and the follow-up paper \cite[][Dinos-II]{sheu25_dinos_ii} acknowledged the possibility of deviations from the power-law lens model at the individual system level. They included $\lambda_\mathrm{int}$ as a hierarchical term and constrained it with single-aperture velocity dispersions, resulting in $\rm \left \langle \lambda_{int} \right \rangle = 0.91^{+0.10}_{-0.09}$ and $\rm \left \langle \lambda_{int} \right \rangle = 0.96\pm0.03$, respectively, each indicating no mean deviation from the power law to $\sim1\sigma$. Dinos-I measured intrinsic scatter on $\lambda_\mathrm{int}$ of $<0.13$, and Dinos-II did not measure intrinsic scatter. Our results confirm these hints from Dinos-I and -II with more detailed kinematics and a more robust treatment of the dynamics on a lens-by-lens basis.

On average, we find nearly isothermal power-law total mass density slopes ($\rho_{\rm tot}\propto r^{-\gamma}$) of $ \left \langle \gamma \right \rangle = 2.04\pm0.02$ with intrinsic scatter of $0.08^{+0.03}_{-0.02}$. This is in agreement to the second decimal with the median result from the lens models of the larger SLACS sample from Dinos-I ($\gamma = 2.04\pm0.04$) as well as $\sim1\sigma$ agreement with the median results from other previous lensing studies of the SLACS sample: $\gamma = 2.075 \pm 0.024 $ from \cite{etherington22}, and $\gamma = 2.08 \pm 0.03 $ from \cite{shajib21}. The sample is on average reasonably well described by a nearly isothermal power-law mass density profile, which continues to be a remarkably powerful tool for lensing and dynamical studies of ETGs as a first approximation. With the sample-level average consistent with no MST, the sample introduces no measurable bias as an external sample for time-delay studies. However, a pure power law is not sufficiently flexible to precisely describe the true mass density profile of each individual galaxy in the joint modeling framework. 

\begin{figure*}
    \centering
    \includegraphics[width=\linewidth]{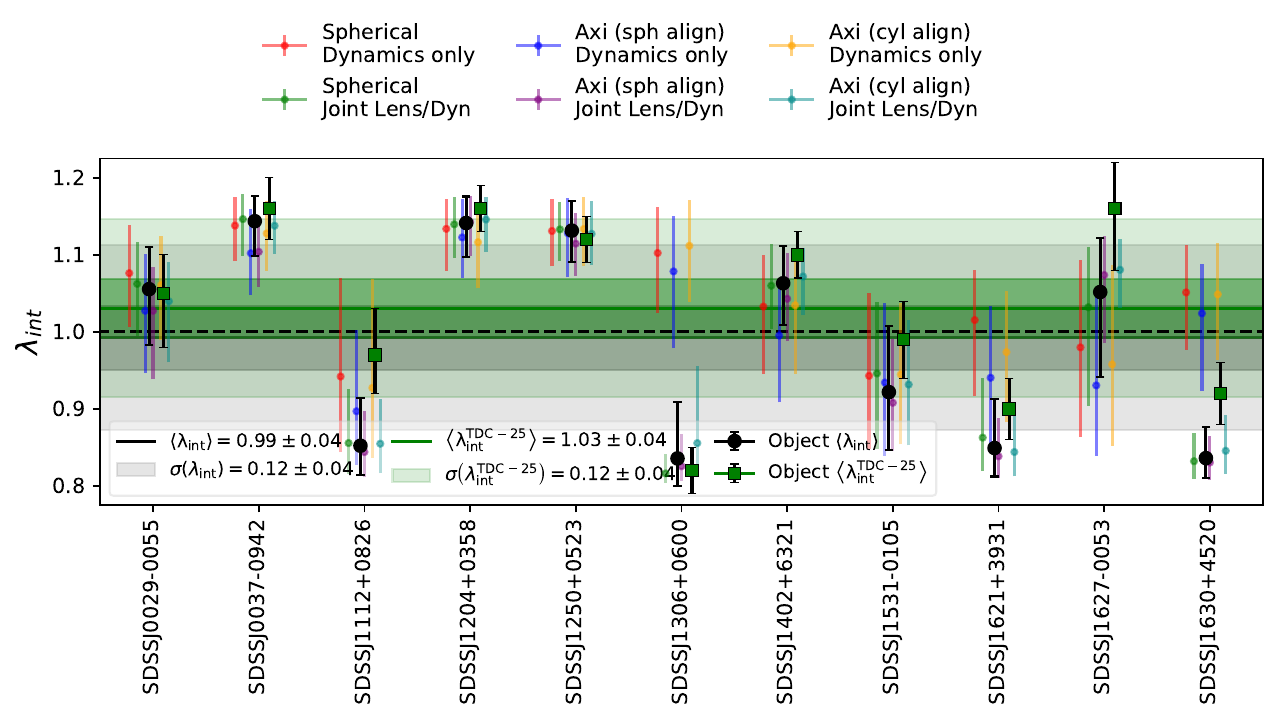}
    \caption{Same as Figure \ref{fig:beta_lambda_plot}. Showing only the objects from TDCOSMO-2025. The inferred best-fit $\rm \lambda_{int}$ from TDCOSMO-2025 is shown in green. }
    \label{fig:beta_lambda_plot_tdc25}
\end{figure*}

Figure~\ref{fig:beta_lambda_plot_tdc25} shows the same plot for only the systems that were included in TDC-25. As in Figure \ref{fig:beta_lambda_plot}, the horizontal lines, darker fill regions, and lighter fill regions show the mean, error on the mean, and scatter of the samples. Green square markers, lines, and fill regions correspond to the best $\rm \lambda_{int}$ values for the SLACS/KCWI sample used in that analysis. Note that the SLACS/KCWI sample in TDC-25 was rebinned to radial shells for the purposes of the hierarchical modeling with spherical Jeans models, and an axisymmetric correction factor from \cite{huang25_triaxiality} was applied afterwards. The $\rm \lambda_{int}$ values we show here are without the axisymmetric correction. Our more complex models with the full kinematic maps agree with the TDC-25 analysis to within $1\sigma$ uncertainties for the sample and on average for the individual lens-to-lens comparison, showing that the scatter reported in TDC-25 remains after relaxing the spherical assumption. Since $\rm \lambda_{int}$ does not depend on the geometry and alignment of the velocity ellipse (see Section \ref{sec:sph_vs_axi}), the switch to axisymmetric modeling does not resolve the scatter. In hierarchical modeling, this scatter will continue to limit the constraining power offered by the SLACS lenses as an auxiliary dataset, even if more systems are added, unless the source of the scatter for this sample can be disentangled and modeled. It is interesting to note that the Einstein radii for SLACS lenses are on average smaller than the effective radii by a factor of 0.57, so the lensing slope is primarily sensitive to a radius well within the effective radius. All kinematic maps in our sample extend to or beyond $\theta_\mathrm{E}$, but only 8/14 reach the effective radius of the object. In this region, the stellar mass dominates, and it is well known that the stellar mass density profile is not exactly an elliptical power law. In contrast, TDC-25 did not detect any intrinsic scatter in $\rm \lambda_{int}$ for the time-delay sample. The TDCOSMO time-delay lenses have higher deflector and source redshifts than SLACS, resulting in Einstein radii that are typically larger than the effective radius. A possible explanation for the difference is that elliptical power laws are a better approximation over this larger radial range than within half the effective radius. Unfortunately the datasets with well determined $\rm \lambda_{int}$ are currently too small to detect any trends with radial extent or any other observable.

\begin{table}[]
    \centering
    \begin{tabular}{l|c|c}
         & $\langle \delta_\mathrm{dyn} \rangle$ & $\langle\delta_\mathrm{joint}\rangle$   \\
         \hline
         $\beta_\mathrm{ani}^\mathrm{sph}$ &  0.141 & 0.138 \\
         $\beta_\mathrm{ani}^\mathrm{cyl}$ &  0.071 & 0.071 \\
         $\gamma$ & 0.050 & 0.046 \\
         $\theta_\mathrm{E}$ & 0.063 & 0.023 \\
         $\lambda_\mathrm{int}$ & 0.082 & 0.058
    \end{tabular}
    \caption{Average errors per object for main parameters of interest, for dynamics-only models and joint lensing/dynamical models. Errors are marginalized over the models for each object using BIC weights, including statistical and systematic errors, except for $\beta_\mathrm{ani}^\mathrm{cyl}$ where we compare only the statistical errors. $\beta_\mathrm{ani}^\mathrm{sph}$ includes spherical and spherically-aligned axisymmetric models. }
    \label{tab:res_errors}
\end{table}

\subsection{BIC model preference}
\label{sec:bic_model_preference}

An example grid of BIC comparisons for one object is shown in Figure \ref{fig:bic_grid}. Here, the first row shows that the spherical model with dynamics only has the lowest BIC, and each element in the row shows the $\rm \Delta BIC$ for the other models, with uncertainties in parentheses. $\rm \Delta BIC \sim \sigma_{\Delta BIC}$ for most of the model comparisons, indicating a distinguishable but mild preference for the best model. We marginalize over these models for each object using object-specific $\Delta \rm BIC$ values for weighting as described by \cite{knabel_mozumdar25_tdcxix}. In the case of the model BICs shown in Figure~\ref{fig:bic_grid}, the best fit model will carry more weight than the others but will not dominate.

Figure~\ref{fig:bic_weights} shows the cumulative stacking of weights for each object of the sample to demonstrate model preferences. The upper panel shows the weights calculated from the $\rm \Delta BIC$s across all 6 models, including the dynamics-only and the joint lensing/dynamical models. Generally, the joint models suffer from lower likelihoods than the dynamical models when being forced to the lens model prior, so they are downweighted when comparing all 6 models. The lower panel shows the weights calculated for only joint lensing/dynamical models. The spherical models are generally preferred over the axisymmetric models for most objects, for the full set of models and for the joint models. The object shown in Figure~\ref{fig:bic_grid} is shown with purple bands in Figure \ref{fig:bic_weights}, where roughly half of the weight of the joint models is carried by the spherical model, which is typical for the sample. For the joint models, 10/14 objects have the highest weight on the spherical models, which carry 45\% of the cumulative weight, while axisymmetric models carry 20\% (spherically aligned) and 35\% (cylindrically aligned).  

The spherical model fits one fewer parameters, so fits that are indistinguishable by the likelihood will slightly prefer the spherical model. However, as displayed in Figure \ref{fig:bic_grid}, all models tend to return comparably successful fits, and the weights are distributed without strong preference. Cylindrically-aligned models are typically better suited to much flatter galaxies with a prominent disk component, rather than the rounder, bulge dominated ETGs in our sample. However, the cylindrically-aligned axisymmetric model is preferred over the spherically-aligned model for 11/14 objects. The reason for this may be due to the extra freedom the cylindrical models have for more radial anisotropy, since they are not constrained by the prior we place on the spherically-aligned anisotropy models, which may give them flexibility to make up for small differences that the power-law mass model fails to capture with  nearly isotropic orbits. We stress that all these differences are small from a statistical point of view.

\subsection{Spherical vs axisymmetric}
\label{sec:sph_vs_axi}

In order to examine the sample in the context of previous TDCOSMO analyses, including TDC-25, we compare the inferred model parameters from our spherical and axisymmetric models. \cite{huang25_triaxiality} showed that spherical JAM models result in a median bias of $\rm \sigma_{sph} / \sigma_{axi} - 1 = 0.017^{+0.007}_{-0.009}$ for the larger SLACS sample \citep{bolton08_slacs_v}, assuming an isothermal profile with isotropic stellar orbits (i.e., $\gamma=2$ and $\beta_\mathrm{ani}=0$). This means that spherical models could bias the inferred velocity dispersion by $1.7\%$, translating to $3.4\%$ in $\rm \lambda_{int}$ and therefore $\rm H_0$. Uncertainties on $\lambda_\mathrm{int}$ for our dataset are on average 5.8\% per object for joint lensing/dynamical models, making it difficult to detect the bias. As shown in Figures~\ref{fig:beta_lambda_plot} and~\ref{fig:beta_lambda_plot_tdc25}, our axisymmetric and spherical models agree within the uncertainties in terms of $\rm \lambda_{int}$, and they agree within uncertainties with $\lambda_\mathrm{int}$ measured for these systems in TDC-25. Direct comparison of the best-fit parameters for each object from axisymmetric and spherical models also shows no detectable bias at the level of uncertainty in the model parameters. 

Given the quality of our data, there is enough freedom for our models to package the differences between spherical and axisymmetric predictions into the residual MSD and MAD space with the parameters $\gamma$, $\beta_\mathrm{ani}$, and $\lambda_\mathrm{int}$, such that the inferred value of $\lambda_\mathrm{int}$ is consistent for spherical and axisymmetric models within the errors. With tighter constraints on the lensing $\gamma$ from better imaging, as is available for several of the time-delay lenses (see Section \ref{sect:priors_bounds}), the uncertainties on $\lambda_\mathrm{int}$ may be small enough to detect the bias. Additionally, the dynamical slopes in our models are not precise enough to overcome the MAD directly, but future dynamical models using JWST-NIRSpec IFS data will also be able to illustrate this more clearly. Larger samples of course would also help to detect any population level effect.

\begin{figure*}
    \centering
    \includegraphics[width=0.9\linewidth]{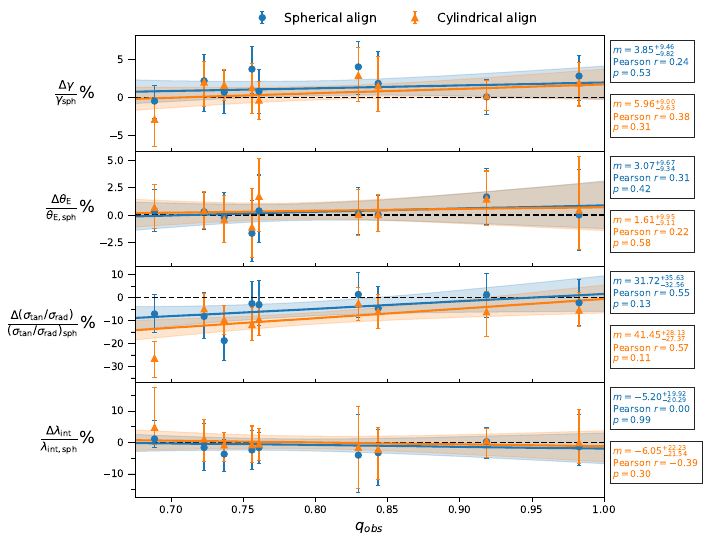}
    \caption{Comparison of axisymmetric and spherical models in terms of mass and anisotropy parameters (slope, Einstein radius, anisotropy ratio, and MST parameter) for joint lensing/dynamical models, given as percent difference with respect to the spherical model result, plotted with respect to the observed axis ratio. Blue points are spherically-aligned axisymmetric models, and orange points are cylindrically-aligned axisymmetric models. Linear fits are calculated with intrinsic scatter and a pivot at the median $q_\mathrm{obs}$ for the sample. Intrinsic scatter is 0 for all fits to within 0.03\% uncertainty. We show the slope, Pearson coefficient, and p-value for each correlation. None of the correlations are significant. See Section \ref{sec:sph_vs_axi} for further discussion.
    }
    \label{fig:axi_vs_sph}
\end{figure*}

\begin{figure}
    \centering
    \includegraphics[width=\linewidth]{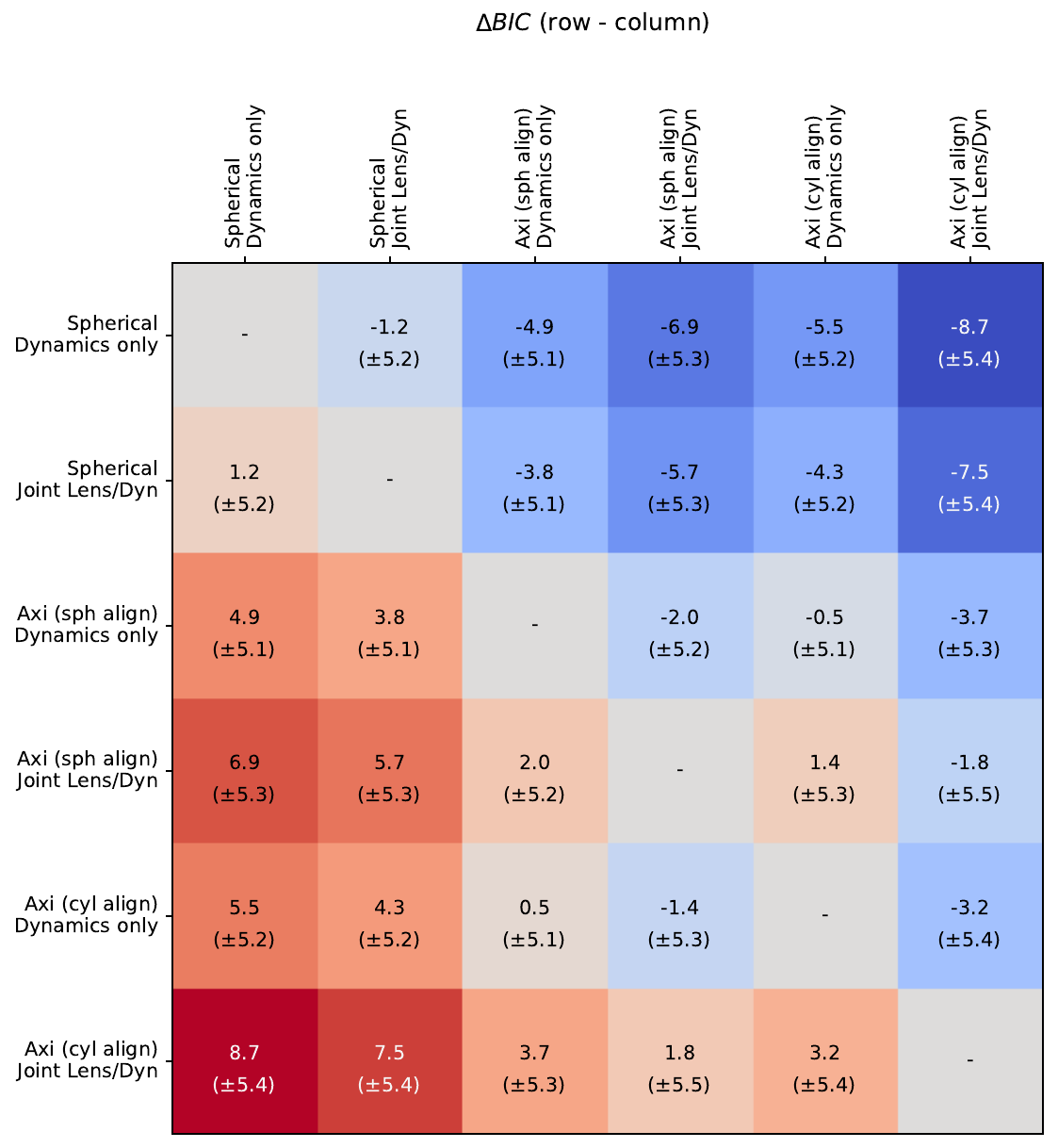}
    \caption{BIC comparisons of models for example object SDSSJ1204+0358.}
    \label{fig:bic_grid}
\end{figure}

\begin{figure}
    \centering
    \includegraphics[width=\linewidth]{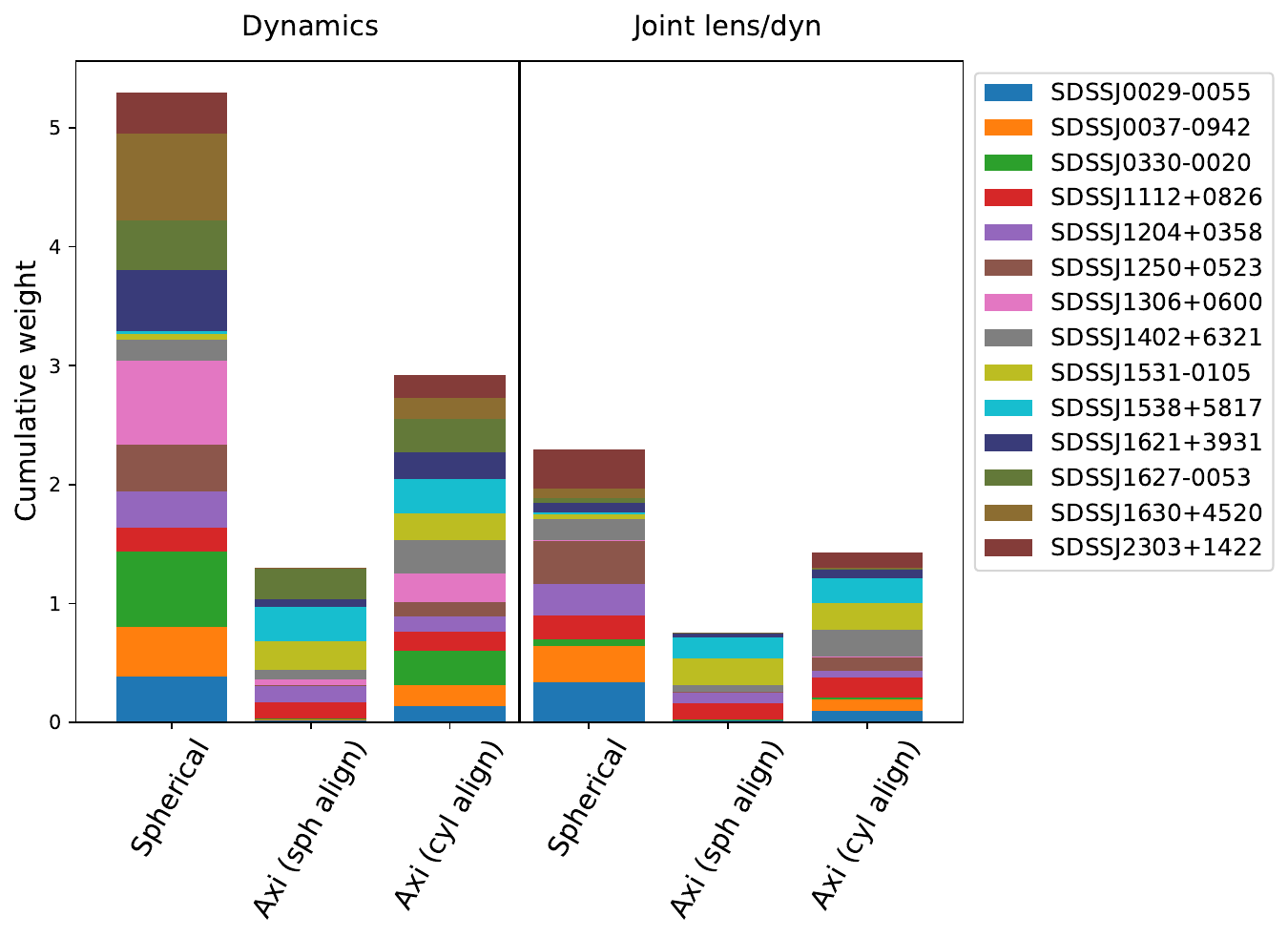}
    \includegraphics[height=0.75\linewidth]{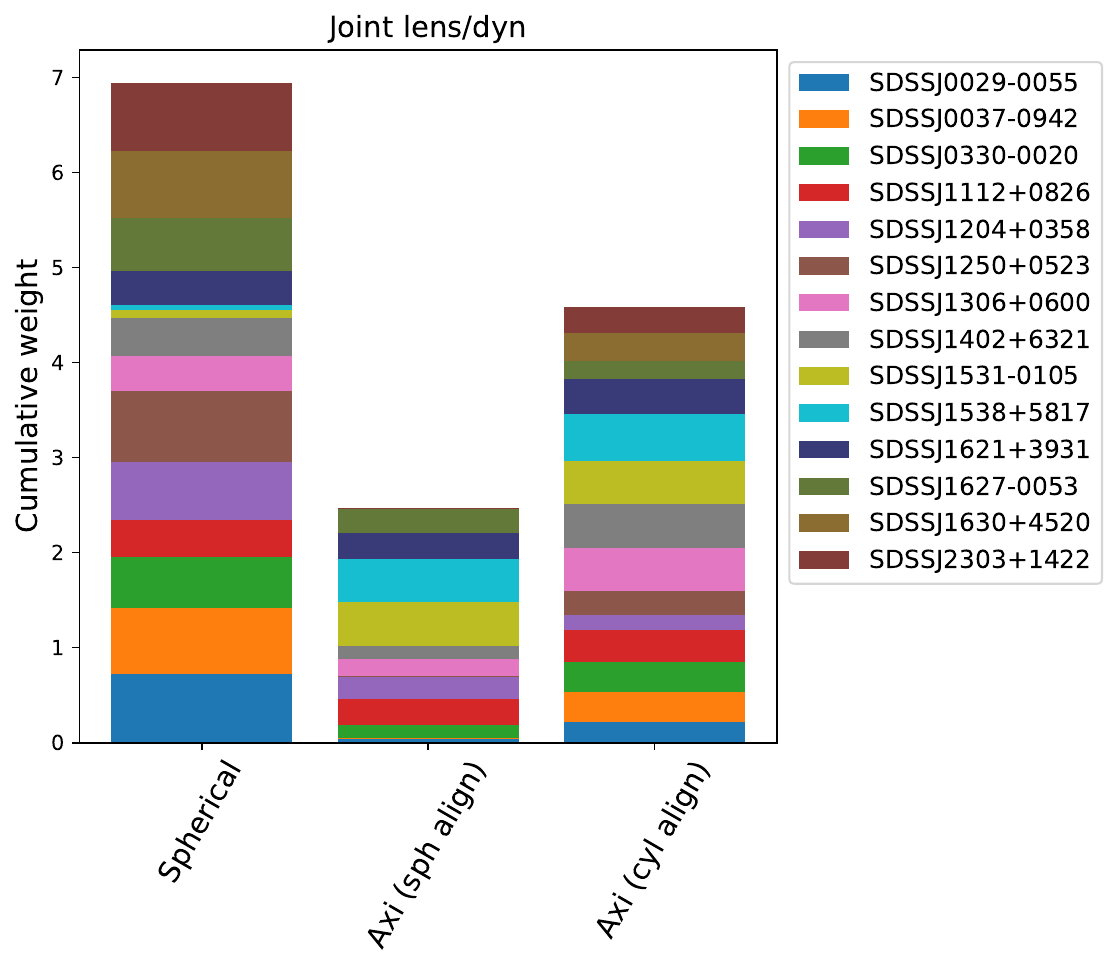}
    \caption{BIC weights for the models for each object.}
    \label{fig:bic_weights}
\end{figure}

Figure~\ref{fig:axi_vs_sph} shows mass and anisotropy parameters for the joint lensing/dynamical models. We compare separately the axisymmetric models with spherical alignment and those with cylindrical alignment. We plot each parameter as a function of observed axis ratio $q_\mathrm{obs}$. For the mass parameters, all data points are consistent with no deviation within the errors, even in cases with the most flattened isophotes, indicating that the bias is not shown with statistical significance. We perform linear fits to each parameter combination with intrinsic scatter and a pivot at the median $q_\mathrm{obs}$ for the sample and find no significant trends with respect to the observed axis ratio for any mass parameter. For each linear fit, we show the linear slope, the Pearson coefficient indicating the strength of correlation, and the p-value. 

The power-law slope and anisotropy parameters are mildly sensitive to the geometric assumptions and ellipticity, and there is a residual degeneracy owing to MAD. The power-law slope is on average $1.3\%\pm1.0$ larger for spherically-aligned axisymmetric models compared with the purely spherical models, and $0.8\%\pm0.9$ for the cylindrically-aligned axisymmetric models, and there is no significant correlation with the axis ratio. The anisotropy shows an offset in the opposite direction, where the anisotropy ratios $\sigma_\mathrm{tan}/\sigma_\mathrm{rad}$ for spherically-aligned axisymmetric models are $5.2\%\pm3.1$ smaller (more radially anisotropic) than the purely spherical models, and $9.0\%\pm2.5$ smaller for cylindrically-aligned axisymmetric models. 

Only anisotropy shows any correlation with the axis ratio. The spherically-aligned case is not significant, as the slope is consistent with 0 within errors. The anisotropy for cylindrically-aligned models is constrained to be less than one (radial) by the prior and is comparable to the spherical case only near the major axis of the ellipsoid, so the correlation shown here is not meaningful. Furthermore, the correlation is strongly driven by the outlying object with the smallest axis ratio; the correlation disappears when this data point is removed. We marginalize over the spherical and axisymmetric models with our BIC weighting scheme as in Figure \ref{fig:beta_lambda_plot}. This is why the average uncertainties on $\beta_\mathrm{ani}$ are larger than the uncertainties for each individual model. 

The parameters that primarily affect the mass normalization ($\rm \theta_E$ and $\rm \lambda_{int}$) are completely consistent between axisymmetric and spherical models and show no correlation with the axis ratio. This is expected and encouraging, since $\rm \lambda_{int}$ is of primary interest to our study. The stability of our models in terms of the mass normalization across the spherical and axisymmetric models confirms the validity of the spherical approach that has been used in previous TDCOSMO papers; however, it reinforces the prudence of utilizing the correction factor introduced by \citep{huang25_triaxiality} when 2D axisymmetric kinematics data is unavailable.  

\subsection{Dynamics-only constraints on $\lambda_\mathrm{int}$}

The main effect of the mass sheet that is detectable to our joint modeling is the overall normalization of the mass profile. For two of the objects in the sample, SDSSJ1306+0600 and SDSSJ1630+4520, the joint lensing/dynamical models differ from the dynamics-only models by more than $1\sigma$, indicating that the lens model $\theta_\mathrm{E}$ is necessary to isolate the effect of the MST for these objects. However, for some of the objects in the SLACS/KCWI sample with higher S/N and better spatial resolution, we find that the dynamical model alone significantly favors $\lambda_\mathrm{int} \neq 1$. Since the overall normalization is a free parameter $\left(\theta_\mathrm{E}\right)$ with a broad uniform prior, it is degenerate with the normalization effect of the mass sheet. The dynamics-only models can measure it from the subtle deviation of the shape of the power-law mass density profile expressed through the mass sheet, which is most noticeable in the outer regions of the galaxy. For this reason, a model with the flexibility to adjust with the MST through the parameter $\lambda_\mathrm{int}$ will be especially sensitive to the outermost spatial bins of the data. These outer regions also tend to have the most contaminating light from the lensed background source arcs. However, with the constraints on the MAD from the anisotropy prior, if the kinematics measured in bins close to and beyond the effective radius are reliable, then they are sensitive to the variations in the mass profile that approximate the MST and can constrain $\lambda_\mathrm{int}$ even without the lens model. We explore whether the outermost bins are accurate enough to constrain this aspect of the mass sheet through dynamical models alone.

The systematics we explored in Paper I were concerned with the spectral fitting method. We showed that covariance between high-S/N spatial bins and samples is at the 1\% level. However, in order to leverage the spatial extent and resolution of the data, we binned the datacubes using Voronoi tessellation with {\sc VorBin} \footnote{We use v3.1.5 of the Python package from \url{https://pypi.org/project/vorbin}.} \citep{capp03_vorbin}. Outer regions of the galaxies are binned with a large number of low-S/N spaxels to reach our target bin S/N of 15 per \AA. We consider the possibility of unknown systematics specific to the low-S/N spaxels to which the bins composed of higher-S/N member spaxels are not subject and which do not average out in the binning process. In addition, it is difficult to quantify the uncertainty introduced by contamination from the background source arcs. As discussed in Paper I, 2 out of 14 objects suffered from enough contamination from the background galaxy flux to noticeably interfere with the Voronoi binning and kinematic extraction steps. In those cases, manual masks were placed over the affected regions before binning. With the difficulty of deblending the foreground deflector light from the background source arcs, all kinematic maps included spaxels with potential source arc contamination.

We test the dependence of our dynamical model inferences on the outer bins by focusing on three objects that appear to prefer $\lambda_\mathrm{int} \neq 1$ (SDSSJ0037-0942, SDSSJ1112+0826, and SDSSJ1204+0358) from dynamical models alone. SDSSJ1112+0826 is consistent with $\lambda_\mathrm{int}=1$ within $1\sigma$ uncertainties, but we include it to examine a galaxy with $\lambda_\mathrm{int}<1$. We inflate the uncertainties of bins where the mean S/N per spaxel $\rm \left < S/N \right>_\mathrm{spax}$ in the bin is low, and where there is significant background source arc contamination. 

For each object, there is a jump in the number of member spaxels and mean S/N per spaxel for bins at roughly 0.7 $R_\mathrm{eff}$, where the mean $\rm \left < S/N \right>_\mathrm{spax} < 4$ for the Voronoi bins. We inflate the total uncertainty of these bins by a multiplicative factor $f_{\rm inf} = 4 \ / \rm \left < S/N \right>_\mathrm{spax}$. 
We identify bins with significant source arc contamination using HST imaging of the objects and B-spline models of the lens deflector light \citep{bolton08_slacs_v}. We chose these models for this exercise instead of the light profiles from Dinos-I for convenience of operation. We subtract the lensing deflector light model from the HST image, and for a well-deblended lens light the residual shows an image of the background source arcs. We convolve with the KCWI PSF (the HST PSF is negligibly small in comparison, so we neglect it for these purposes), interpolate over the HST grid, and map to the rotated KCWI grid. From this source arc image, matched to the KCWI datacube, we identify all bins that contain pixels with source arc flux above $3\sigma$ from the background level, masking the central core where the deflector light model overpredicts the luminosity profile of the data. We make another map of the relative surface brightness of the source arcs compared to the light of the deflector model, and we identify bins containing spaxels with source arc brightness equal to or greater than the deflector light in that spaxel. All bins identified this way are contaminated by source light, and we flag them for our tests. We add a 5\% systematic to all spatial bins contaminated by source arcs; bins with $\rm \left < S/N \right>_\mathrm{spax} < 4$ and source-arc contamination first have the 5\% systematic added in quadrature and are then rescaled according to $\rm \left < S/N \right>_\mathrm{spax}$. We also test the effect of removing all bins identified this way from the fits.

As expected, inflating the bin uncertainties increases the overall uncertainties for the inferred mass model and anisotropy parameters. However, the mean does not shift. For models that prefer $\lambda_\mathrm{int} \neq 1$, the inflated uncertainties still prefer the same values. This is true for $\lambda_\mathrm{int} > 1$ and $\lambda_\mathrm{int} < 1$, and the lens model information tightens constraints on all parameters. When we remove the affected outer bins from the fits, the dynamics-only models do not constrain $\lambda_\mathrm{int}$ and exhibit broad distributions centered around $\lambda_\mathrm{int} \sim 1$, indicating that the model is able to fit the central bins adequately by rescaling the normalization without adjusting the power-law shape. However, the joint model is constrained to be close to $\theta_\mathrm{E}$ from the lens model, and the normalization of the model $V_\mathrm{rms}$ requires $\lambda_\mathrm{int} \neq 1$, in perfect agreement with the models that include the outer bins. Together, this supports the inclusion of the spaxels in the outer regions of the galaxy and suggests there is no reason to suspect that there are unaccounted for systematics biasing the final results. The uncertainties adequately capture the quality of those bins' kinematic measurements. Furthermore, visual inspection of the combined spectra in the flagged bins showed no specific evidence of defects or contaminating emission lines from the background source arcs. We conclude that our kinematic maps are high enough quality for our dynamical models to be sensitive to this aspect of the internal MST, given the constraining prior on the anisotropy.

\section{Conclusions}
\label{sec:conclusions}

Our main conclusions are:

\begin{enumerate}
    \item Joint lensing/dynamical models constrain the MSD with average errors (statistical and systematic) on $\rm \lambda_{int}$ of 5.8\% per lens.
    \item The sample has a mean MST parameter of $\rm \left \langle \lambda_{int} \right \rangle = 1.01\pm0.03$ and power-law slope of $\rm \left \langle \gamma \right \rangle = 2.04\pm0.02$, indicating that on average a nearly isothermal power law will describe the mass density profiles of SLACS lenses sufficiently for most galaxy evolution studies, and that it introduces no measurable bias as an external sample for time-delay studies. However, the intrinsic scatter of the sample of $0.11\pm0.03$ in terms of $\lambda_\mathrm{int}$ and $0.08^{+0.03}_{-0.02}$ in $\gamma$ suggest that the individual galaxies may be better described by more complex mass models, perhaps enabling one to extract more information and improve the precision, if one intends to use them for precision cosmology.
    \item The sample is well-fit by both spherical and axisymmetric Jeans models, showing no detectable bias within the uncertainties on $\lambda_\mathrm{int}$. These uncertainties are larger than the median percent-level bias predicted by \cite{huang25_triaxiality}, reinforcing the prudence of utilizing their derived correction factor in joint lensing/dynamical analyses under the assumption of spherical symmetry, as in TDC-25.
    \item Dynamics-only constraints on $\rm \lambda_{int}$ with average errors of 8.2\% show that the kinematics data are capable of detecting deviations from the pure power-law mass model. This suggests that a sample of $\sim70$ non-lensing galaxies  with spatially resolved kinematics, such as the $\sim200$ studied in \cite{mozumdar26_magnus3}, could explicitly constrain $\lambda_\mathrm{int}$ to the $1\%$ level and identify a potential redshift dependence.
    \item While the mass normalization and $\rm \lambda_{int}$ are well constrained by the joint lensing/dynamical models regardless of anisotropy, an informative prior is crucial for constraining the mass profile slope; without it, even the joint lensing/dynamical models do not strongly constrain the anisotropy and thus the slope through MAD. With tighter constraints on the MSD-invariant lensing slope, such as those achieved by lens models for the time-delay sample by the TDCOSMO collaboration, the joint analysis may provide tighter constraints on the MAD independently of the prior we utilized here.

\end{enumerate}

Future work with similar datasets should explore more flexible mass profiles, utilizing the stellar tracer profile with a mass-to-light ratio and a separate dark matter profile to decouple the stellar mass from the dark matter halo. This will be necessary to better understand the scatter on $\rm \lambda_{int}$ for this sample. Otherwise, it remains unclear how representative (and therefore useful) this sample is in hierarchical analysis of strong lensing systems for cosmology. All datasets at these redshifts, and those that include strong lensing, are not of sufficient spatial resolution and signal-to-noise ratio to warrant more complex dynamical models than those attained by axisymmetric Jeans modeling. In the meantime, the anisotropy prior needs to be tested for datasets that can independently break the mass-anisotropy degeneracy. The key will be to continue to observationally bridge the gap in ETG evolution between the redshifts of the time-delay lenses and the nearby galaxies for which our dynamical models are robustly understood.


\begin{acknowledgments}

We thank Pritom Mozumdar for helpful discussion and feedback during the writing of the manuscript. Some of the data presented herein were obtained at Keck Observatory, which is a private 501(c)3 non-profit organization operated as a scientific partnership among the California Institute of Technology, the University of California, and the National Aeronautics and Space Administration. The Observatory was made possible by the generous financial support of the W.~M.~Keck Foundation. 
The authors wish to recognize and acknowledge the very significant cultural role and reverence that the summit of Maunakea has always had within the Native Hawaiian community. We are most fortunate to have the opportunity to conduct observations from this mountain.
We acknowledge support by NSF through grant NSF-AST-2407277, and from the Moore Foundation through grant 8548.

\end{acknowledgments}

\appendix


\section{Models}
\label{sec:corner_plots}

\begin{figure*}
    \centering
    \includegraphics[width=\linewidth]{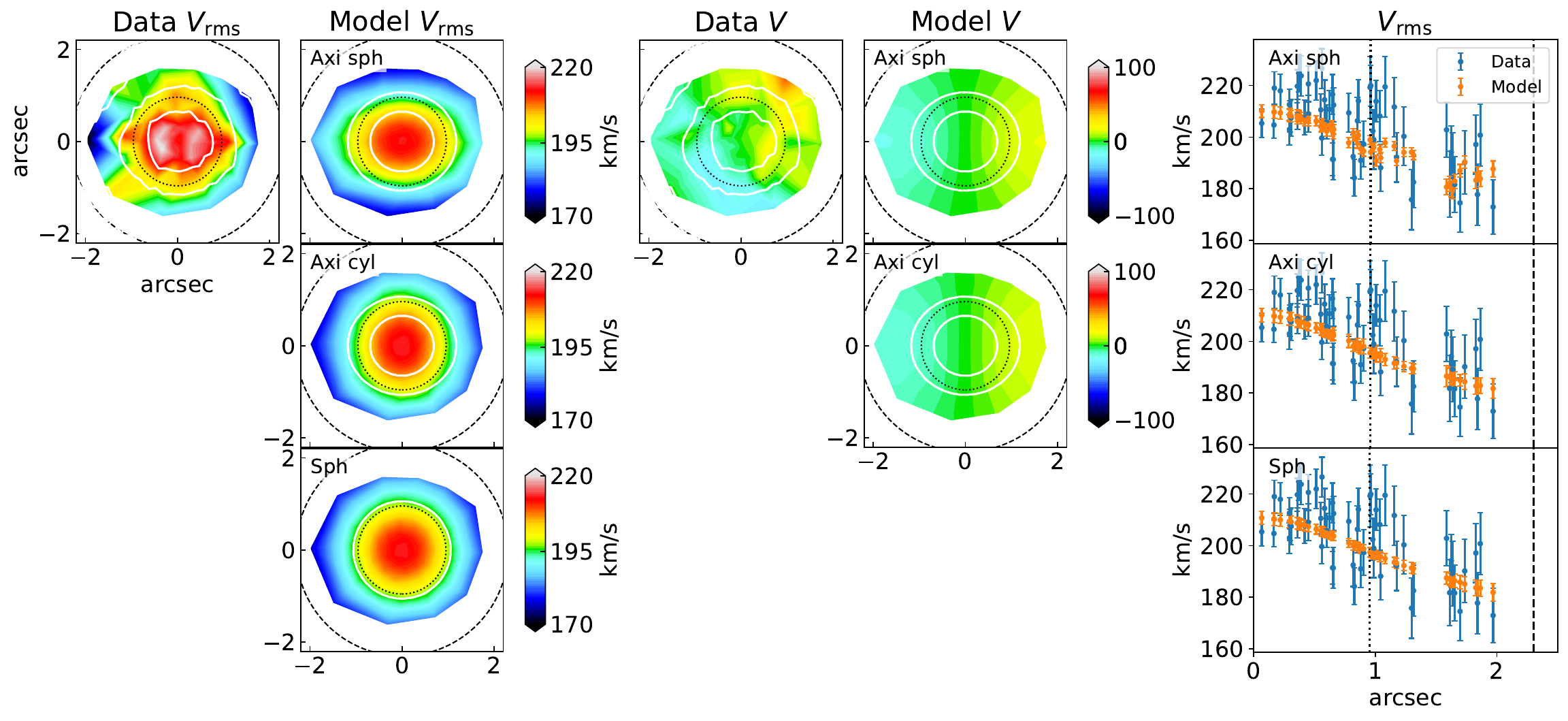}
    \includegraphics[width=0.85\linewidth]{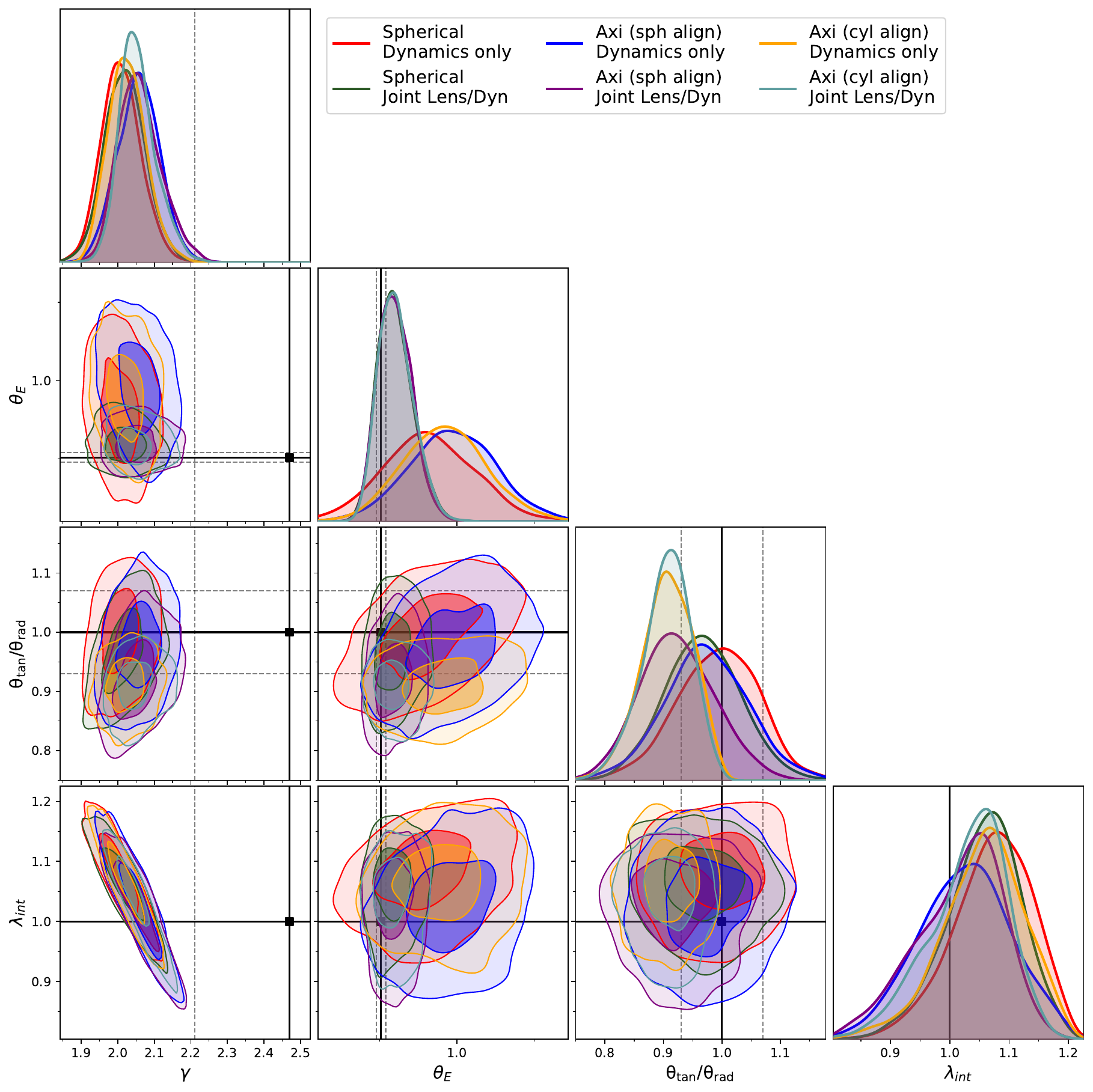}
    \caption{SDSSJ0029-0055}
    \label{fig:j0029}
\end{figure*}

\begin{figure*}
    \centering
    \includegraphics[width=\linewidth]{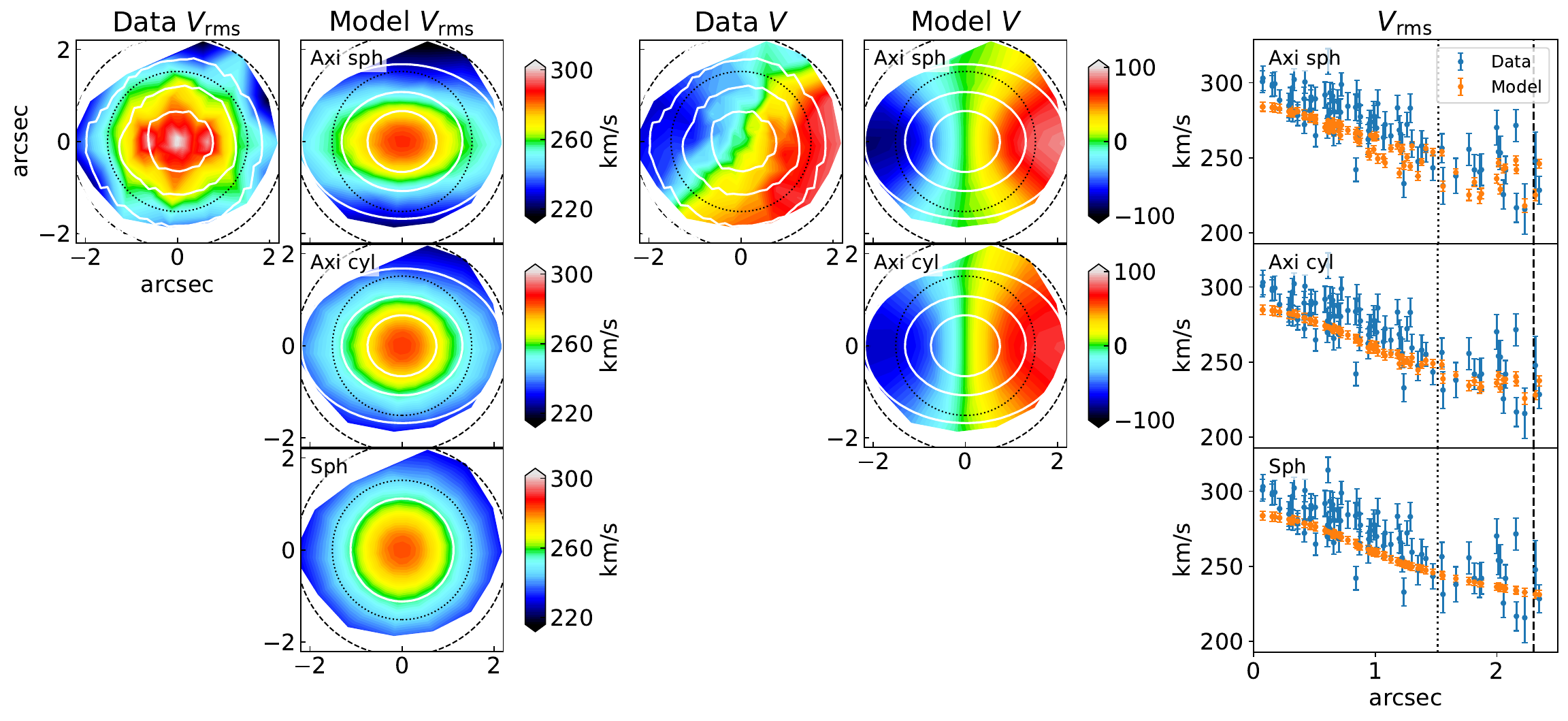}
    \includegraphics[width=0.85\linewidth]{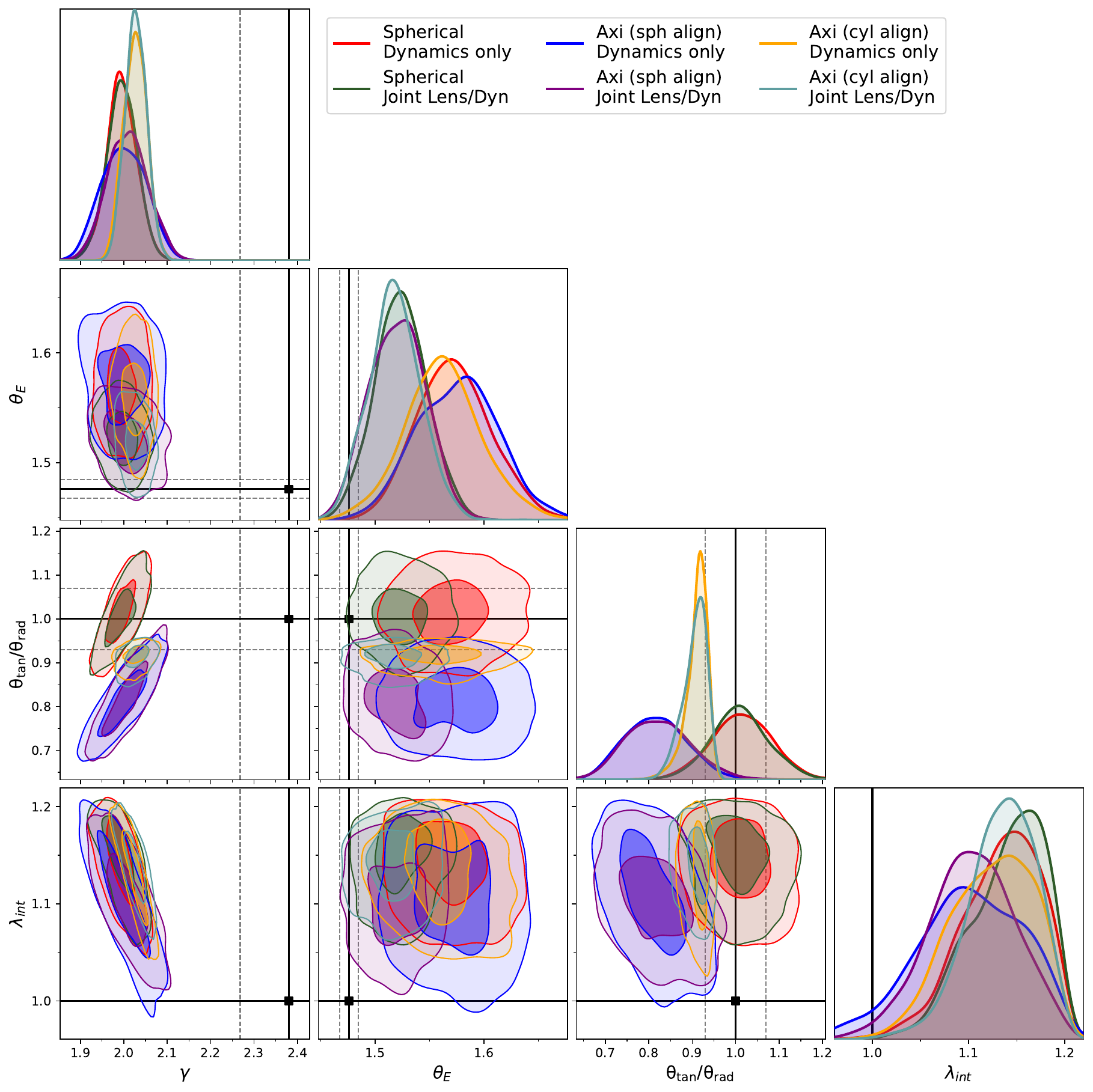}
    \caption{SDSSJ0037-0942}
    \label{fig:j0037}
\end{figure*}

\begin{figure*}
    \centering
    \includegraphics[width=\linewidth]{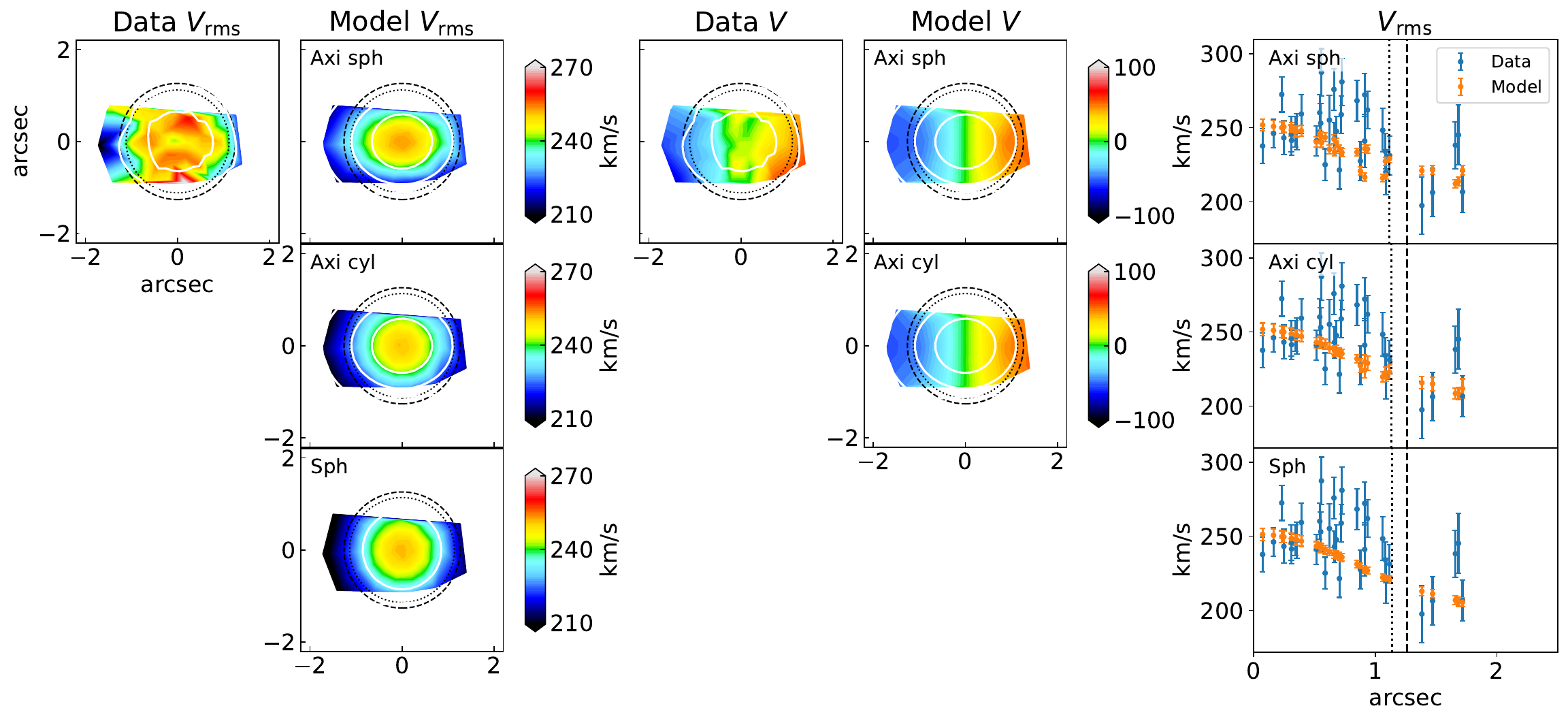}
    \includegraphics[width=0.85\linewidth]{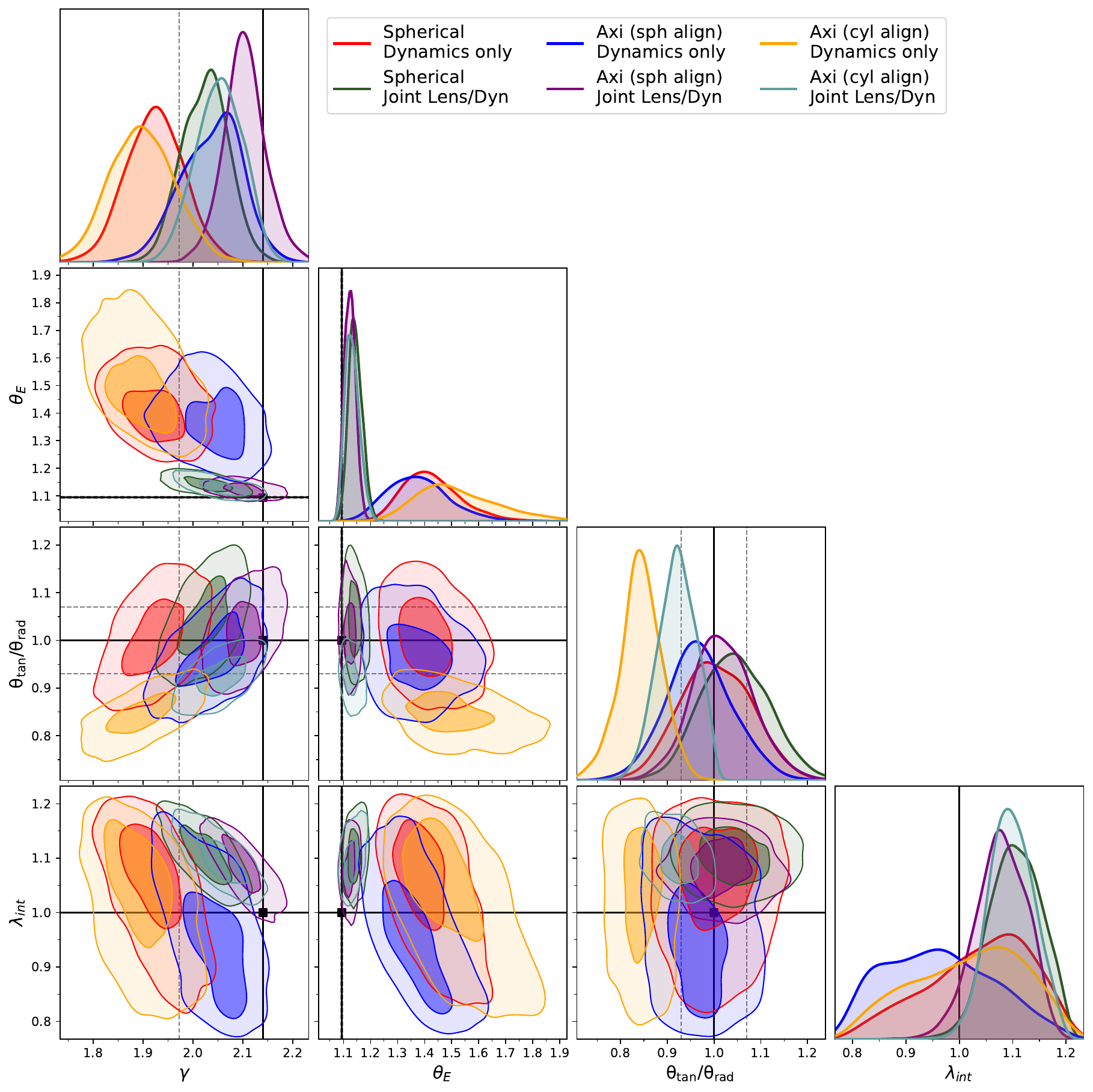}
    \caption{SDSSJ0330-0020}
    \label{fig:j0330}
\end{figure*}

\begin{figure*}
    \centering
    \includegraphics[width=\linewidth]{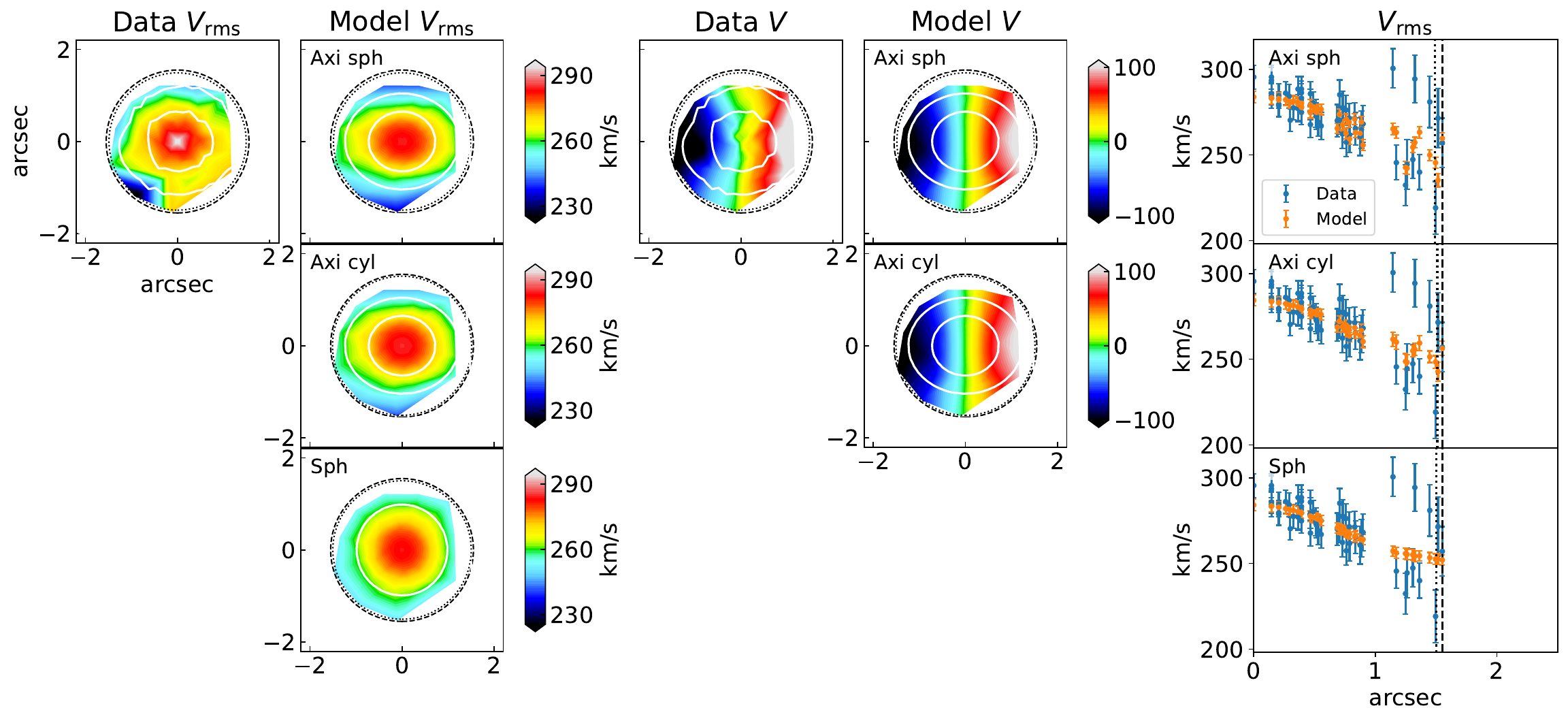}
    \includegraphics[width=0.85\linewidth]{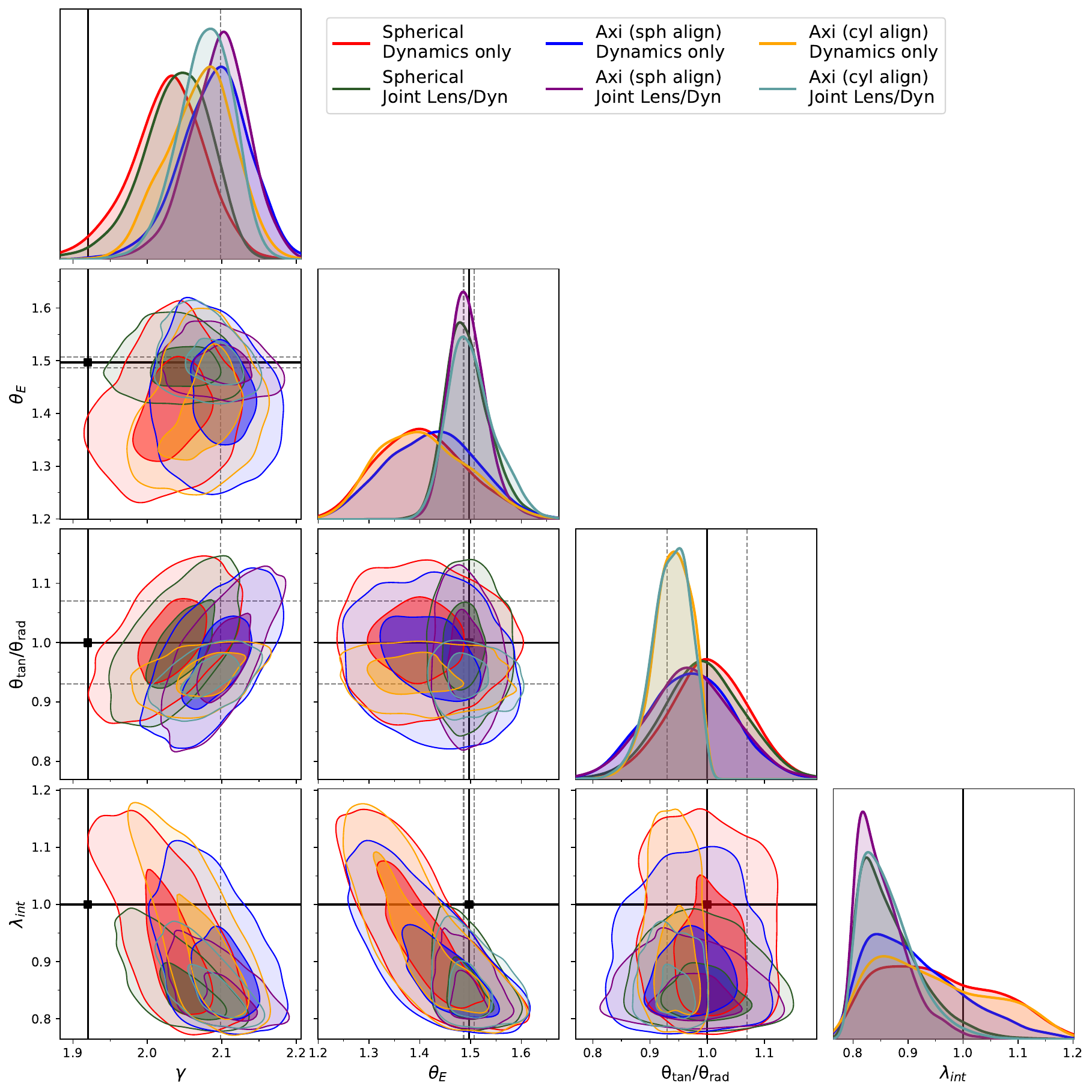}
    \caption{SDSSJ1112+0826}
    \label{fig:j1112}
\end{figure*}

\begin{figure*}
    \centering
    \includegraphics[width=\linewidth]{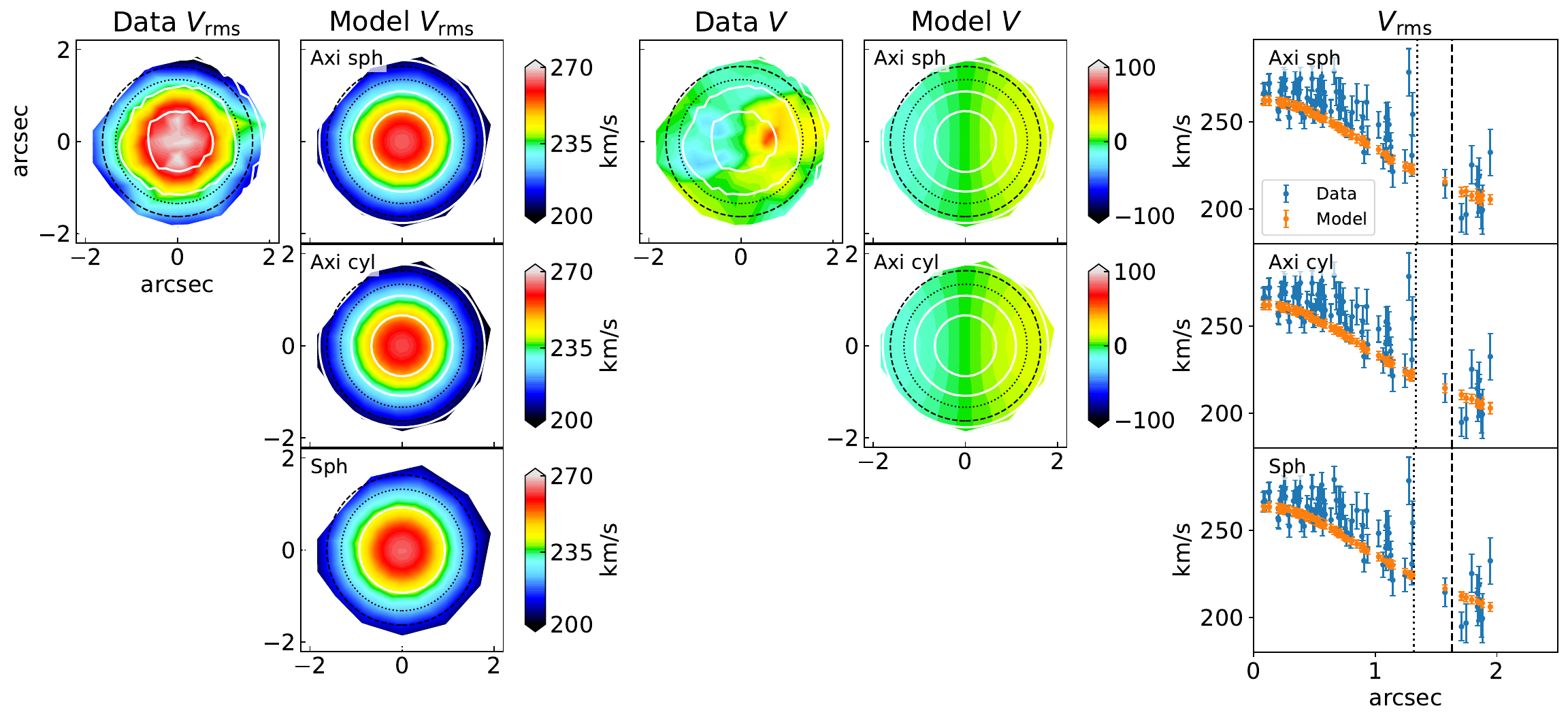}
    \includegraphics[width=0.85\linewidth]{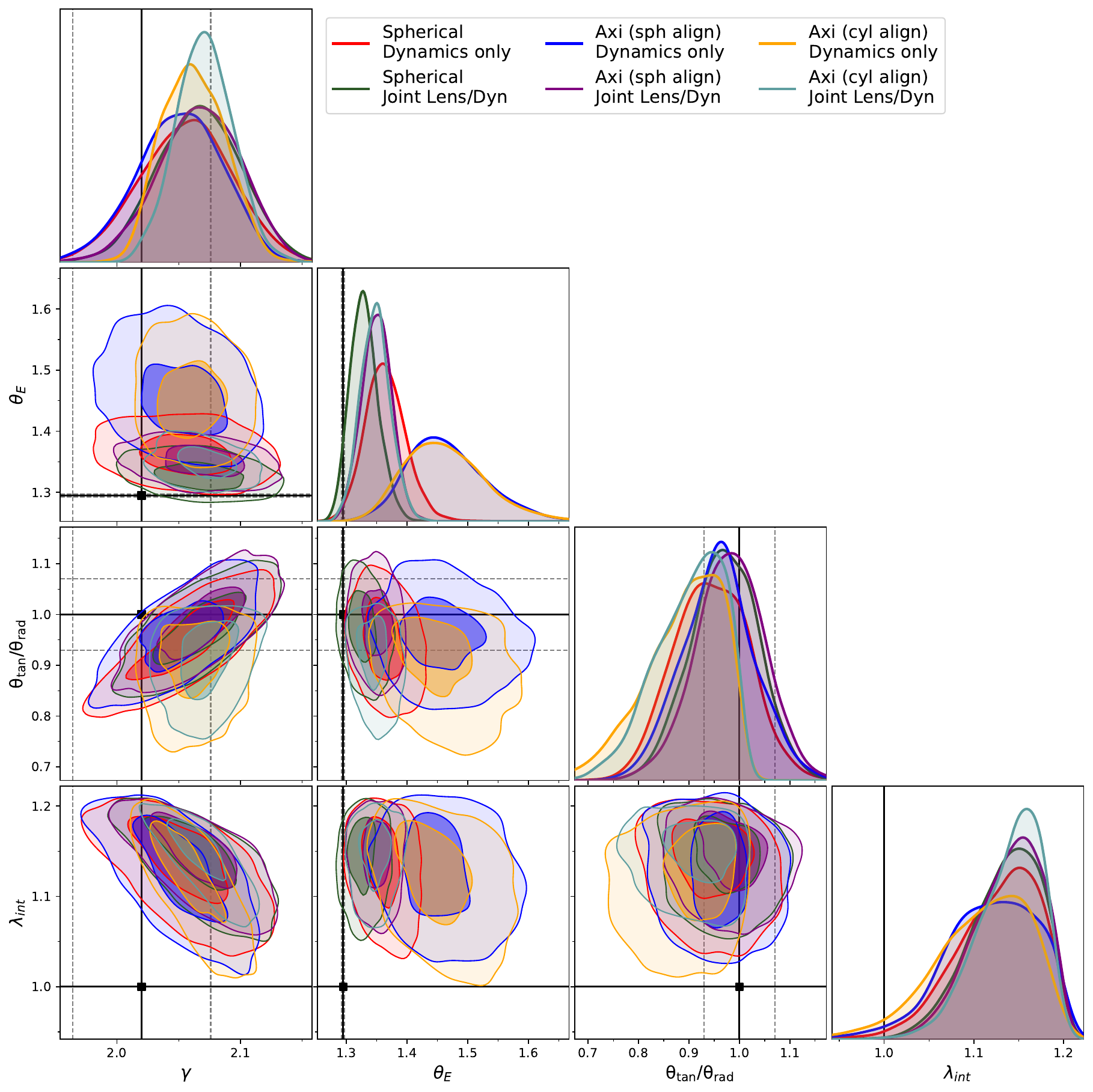}
    \caption{SDSSJ1204+0358}
    \label{fig:j1204}
\end{figure*}

\begin{figure*}
    \centering
    \includegraphics[width=\linewidth]{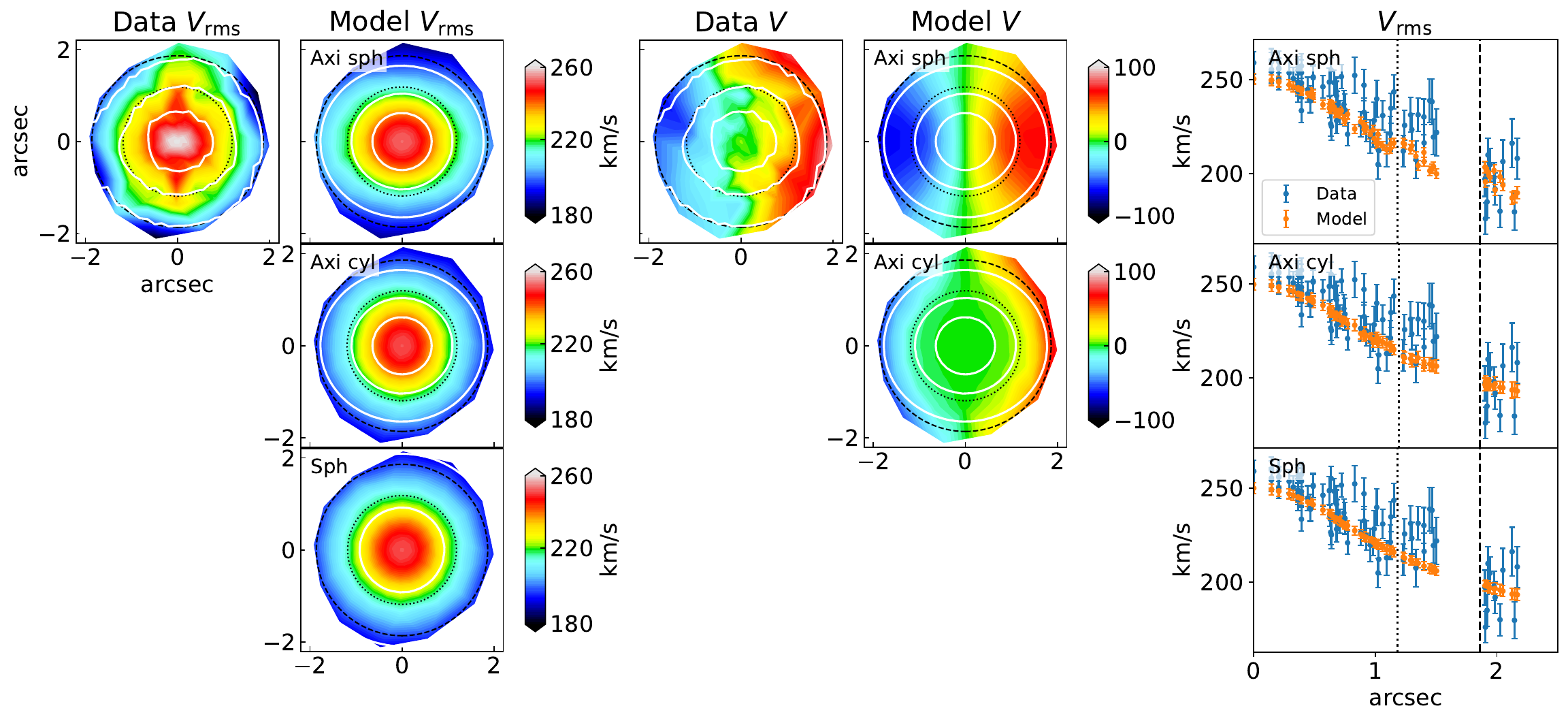}
    \includegraphics[width=0.85\linewidth]{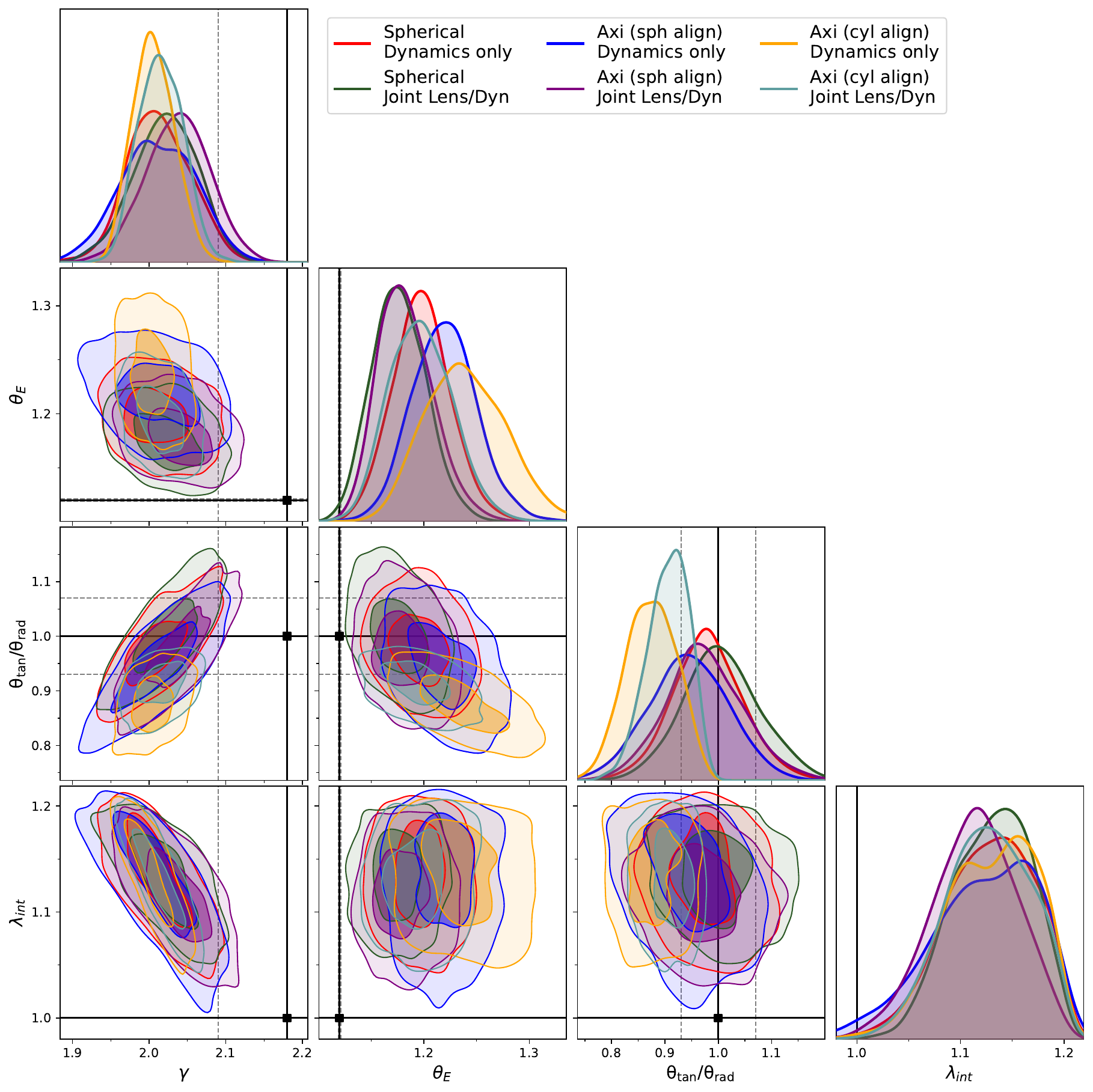}
    \caption{SDSSJ1250+0523}
    \label{fig:j1250}
\end{figure*}

\begin{figure*}
    \centering
    \includegraphics[width=\linewidth]{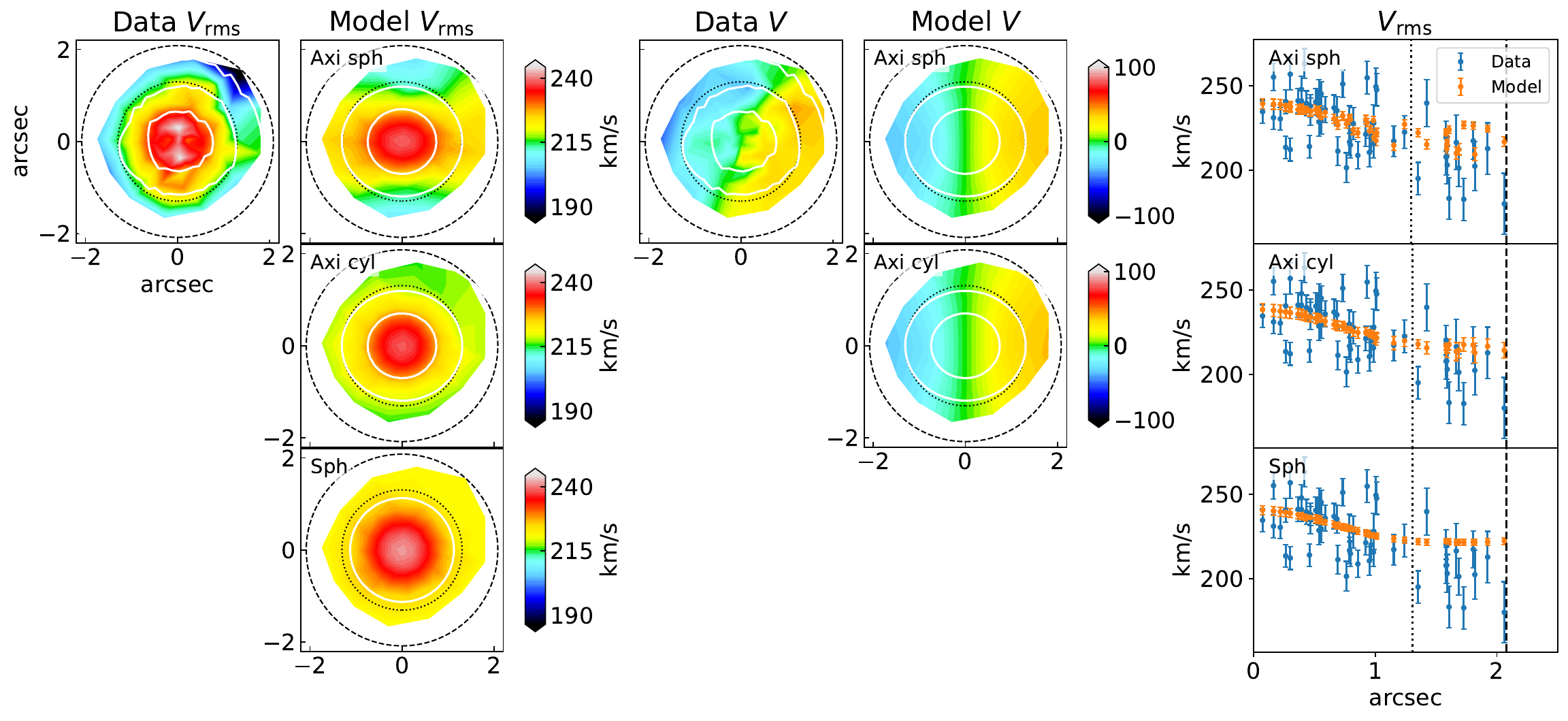}
    \includegraphics[width=0.85\linewidth]{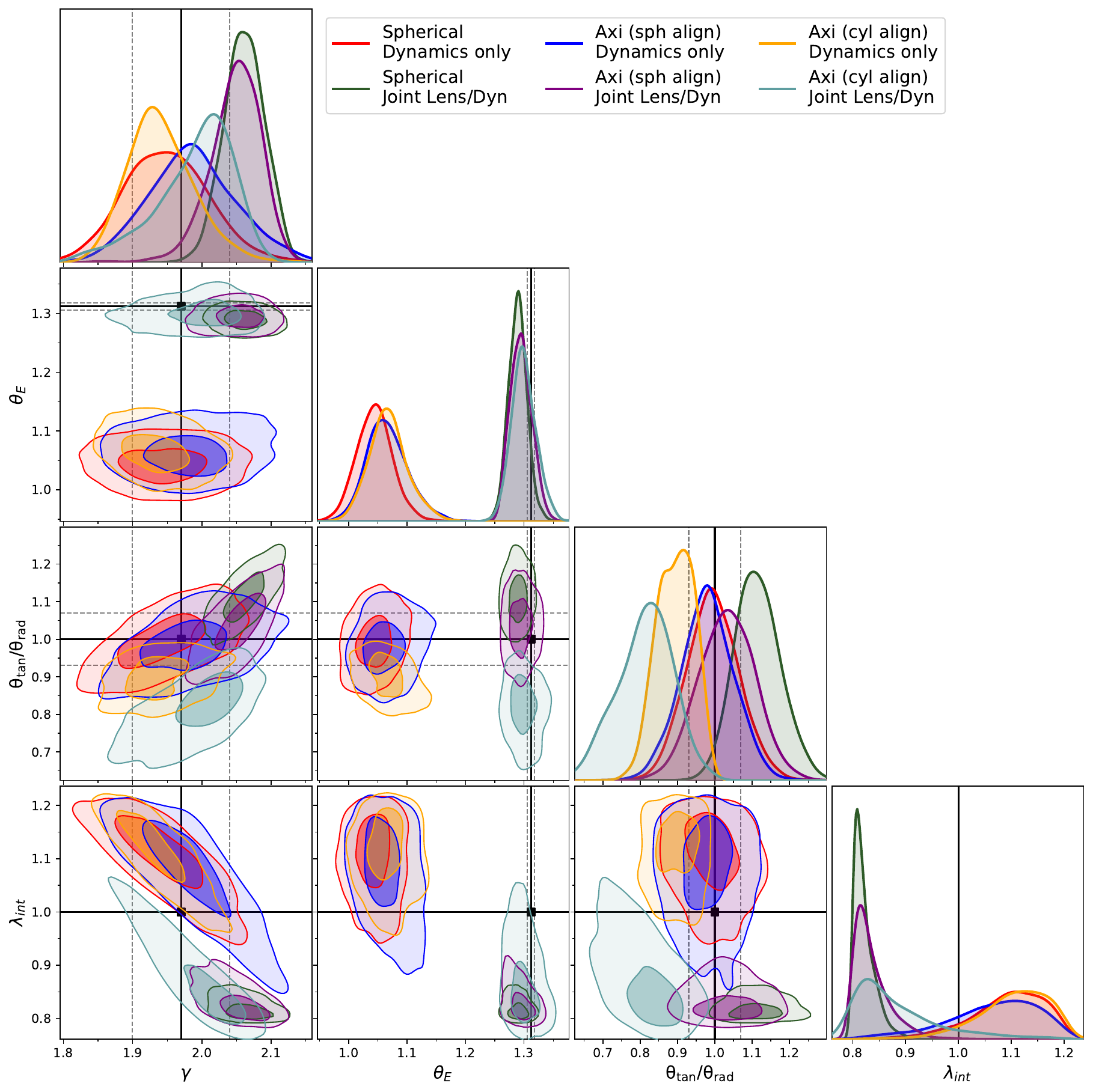}
    \caption{SDSSJ1306+0600}
    \label{fig:j1306}
\end{figure*}

\begin{figure*}
    \centering
    \includegraphics[width=\linewidth]{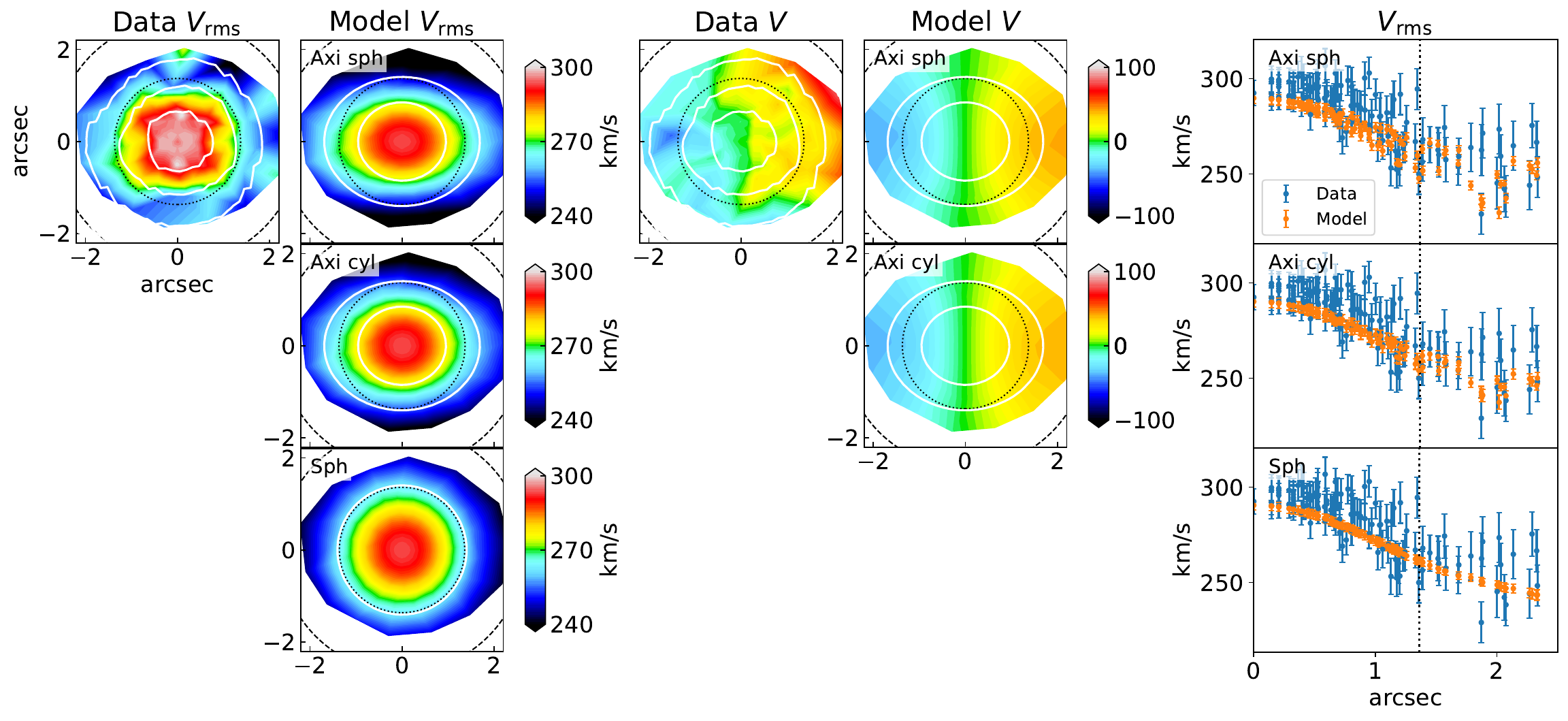}
    \includegraphics[width=0.85\linewidth]{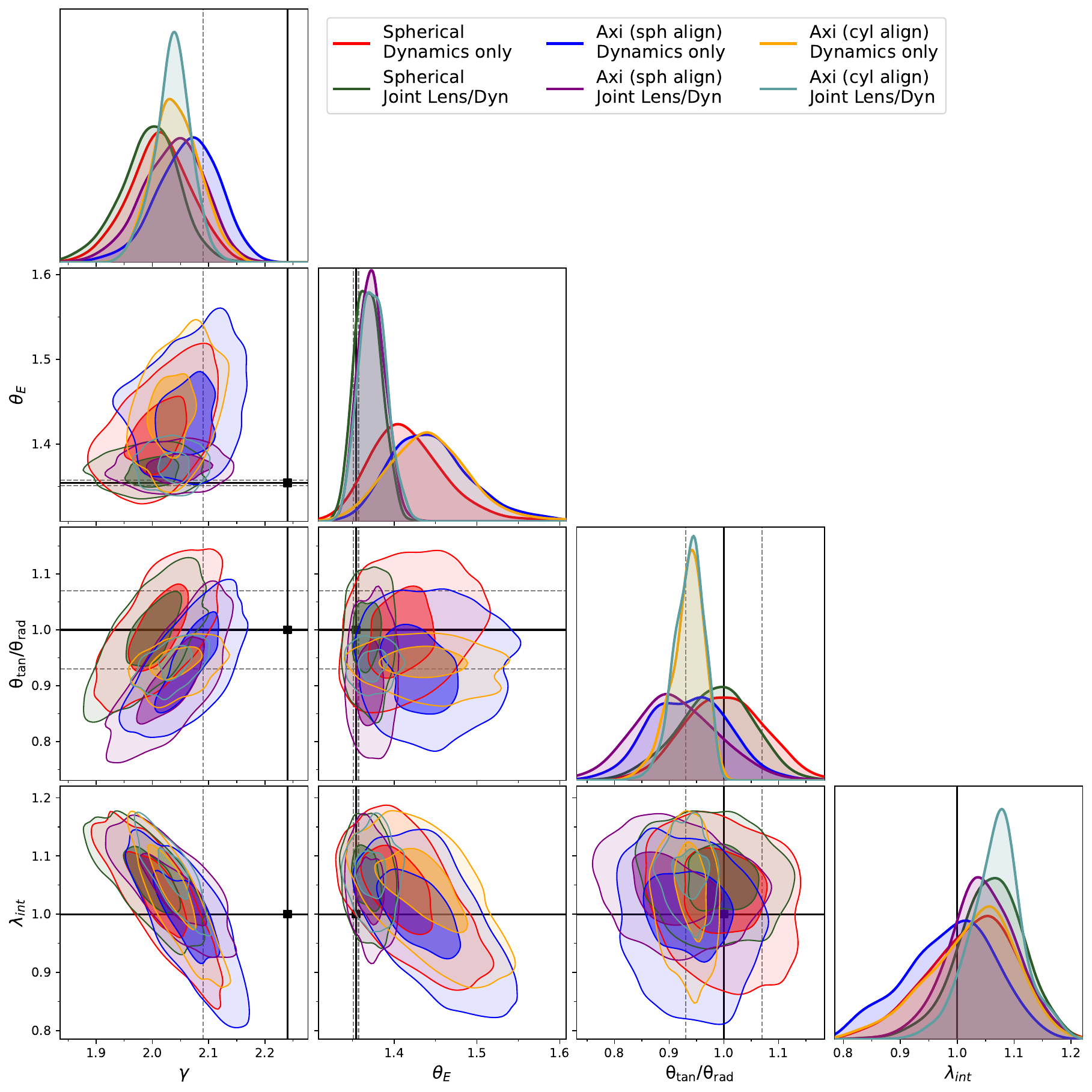}
    \caption{SDSSJ1402+6321}
    \label{fig:j1402}
\end{figure*}

\begin{figure*}
    \centering
    \includegraphics[width=\linewidth]{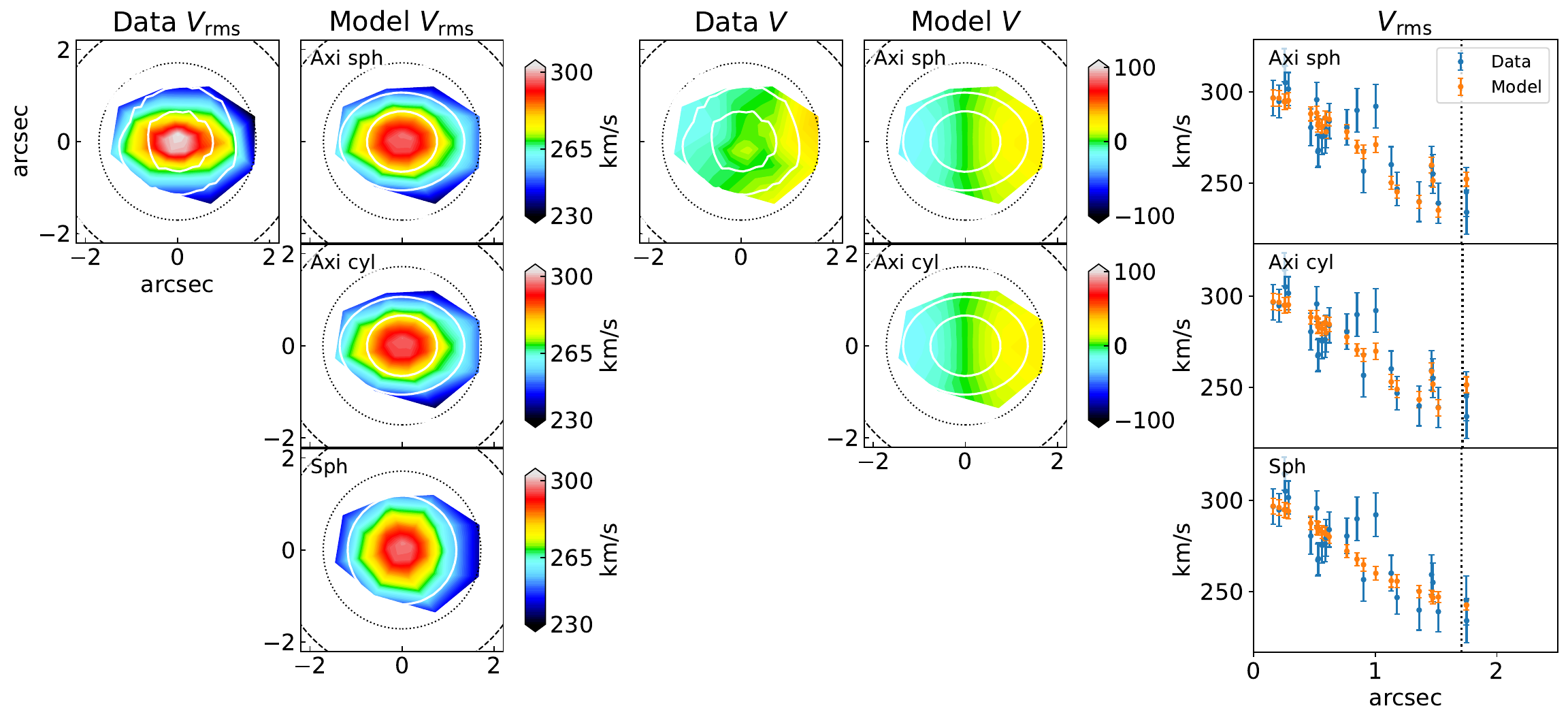}
    \includegraphics[width=0.85\linewidth]{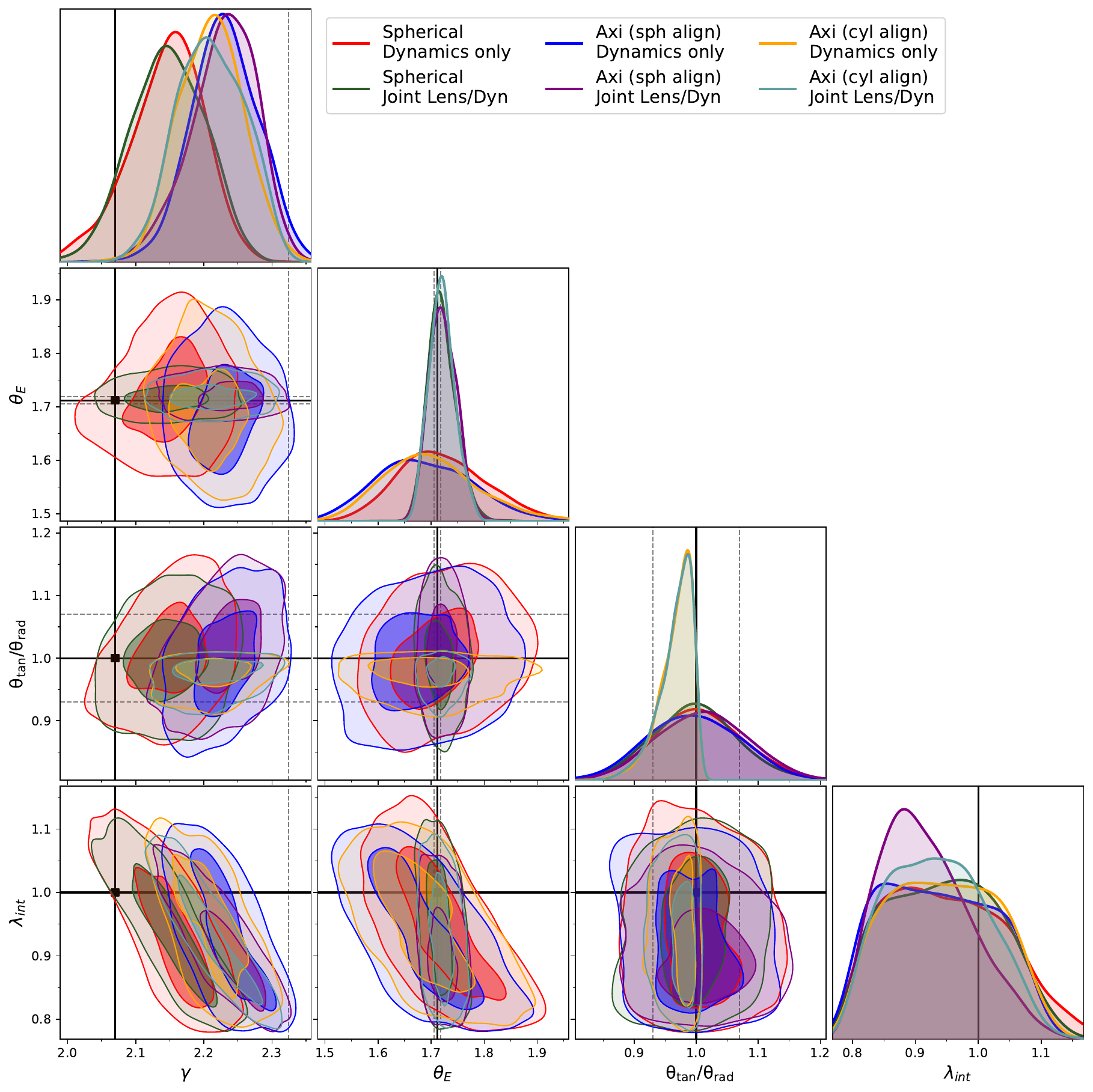}
    \caption{SDSSJ1531-0105}
    \label{fig:j1531}
\end{figure*}

\begin{figure*}
    \centering
    \includegraphics[width=\linewidth]{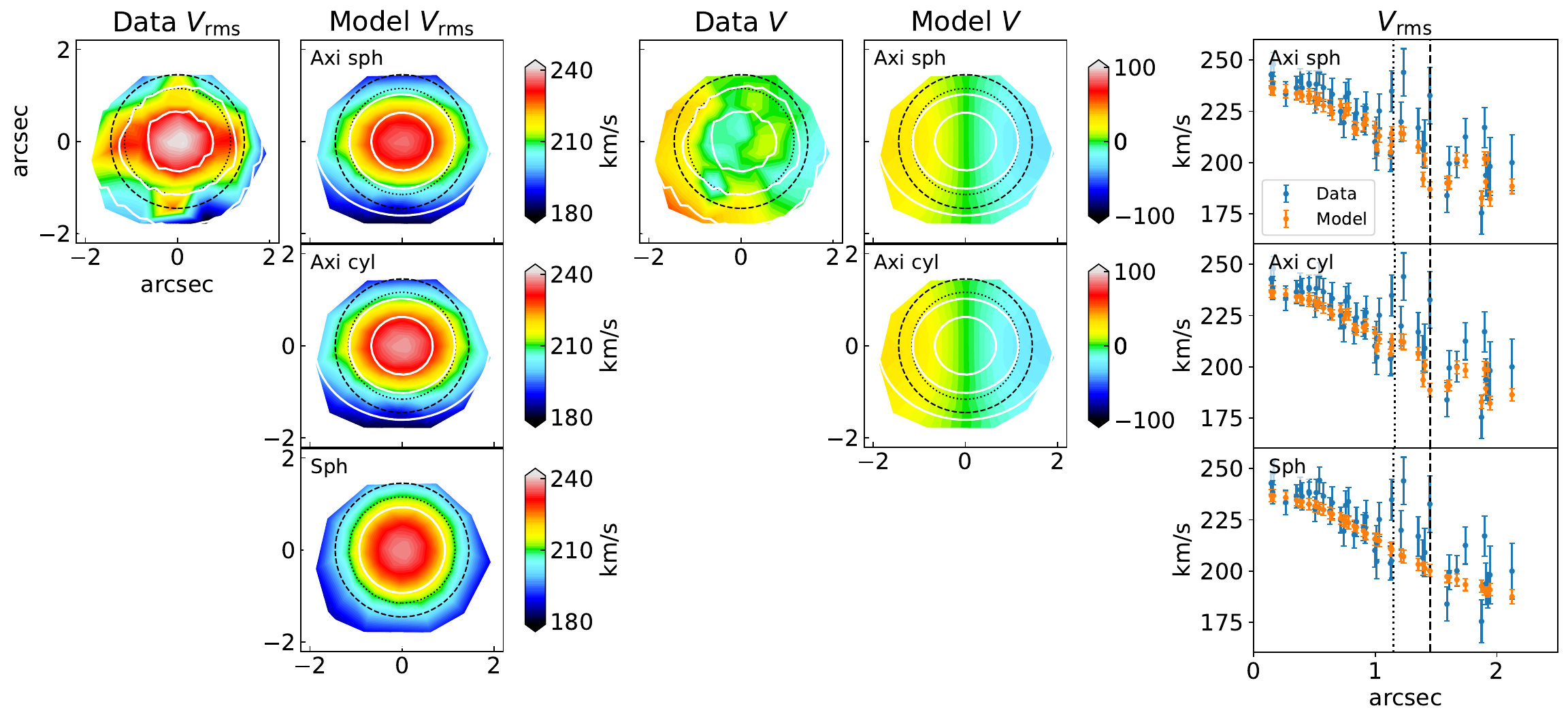}
    \includegraphics[width=0.85\linewidth]{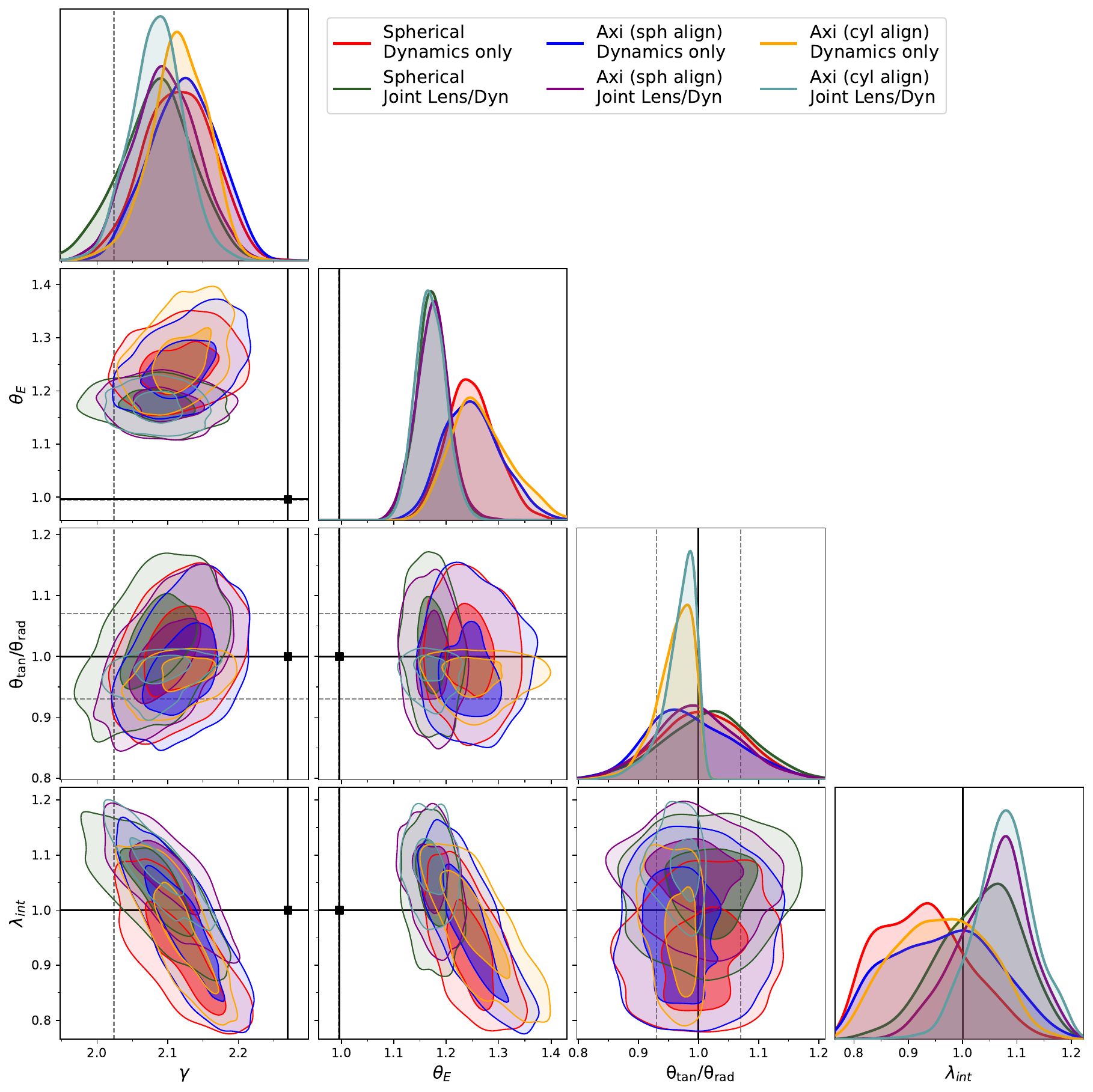}
    \caption{SDSSJ1538+5817}
    \label{fig:j1538}
\end{figure*}

\begin{figure*}
    \centering
    \includegraphics[width=\linewidth]{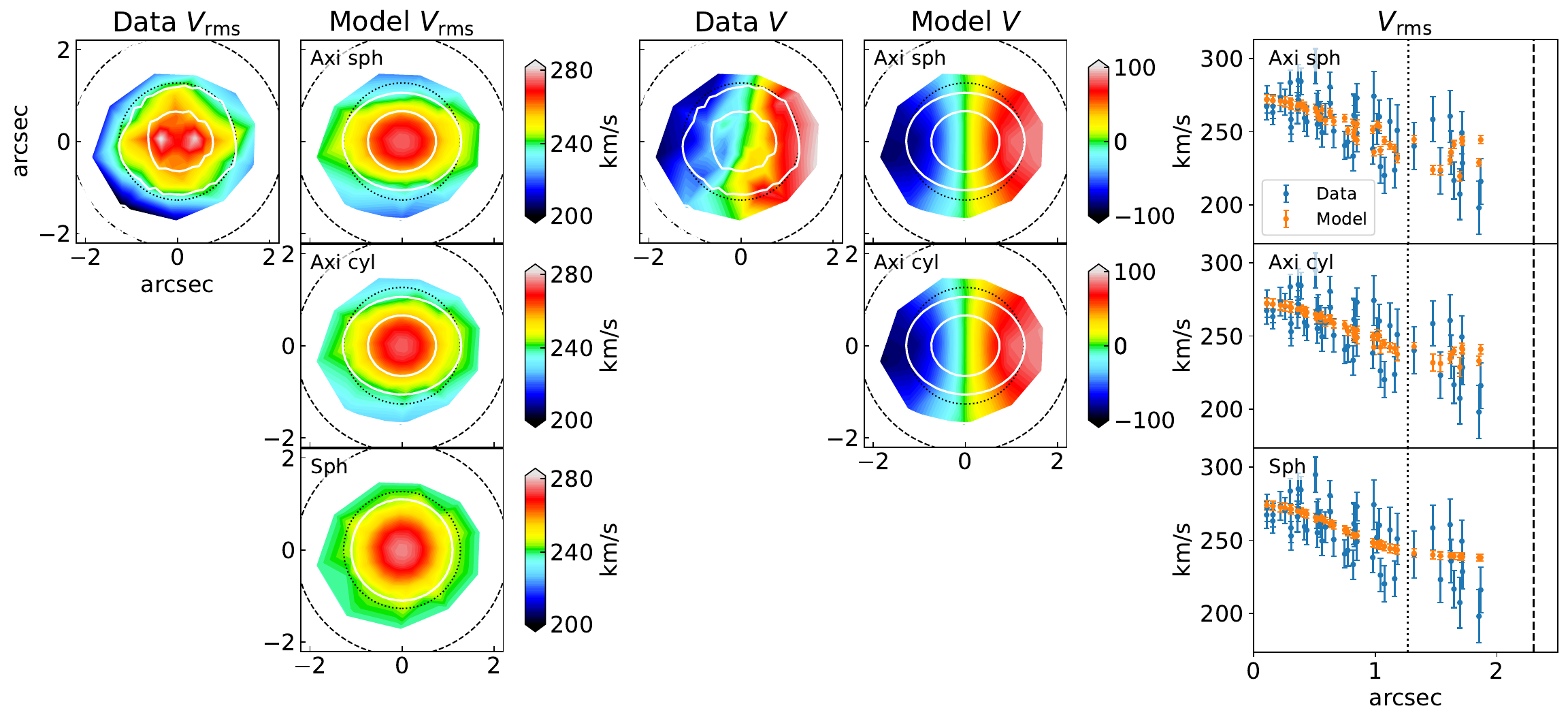}
    \includegraphics[width=0.85\linewidth]{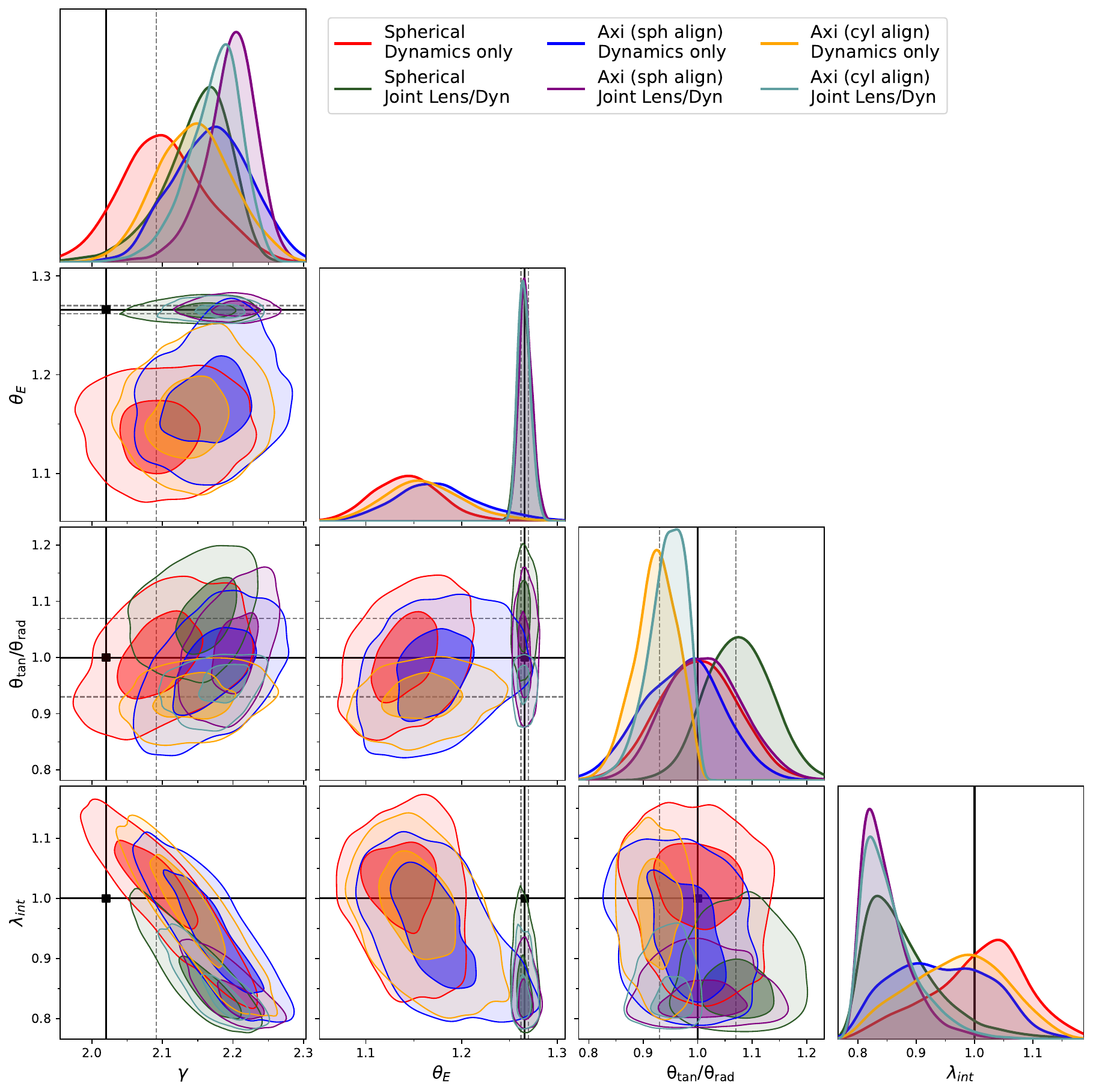}
    \caption{SDSSJ1621+3931}
    \label{fig:j1621}
\end{figure*}

\begin{figure*}
    \centering
    \includegraphics[width=\linewidth]{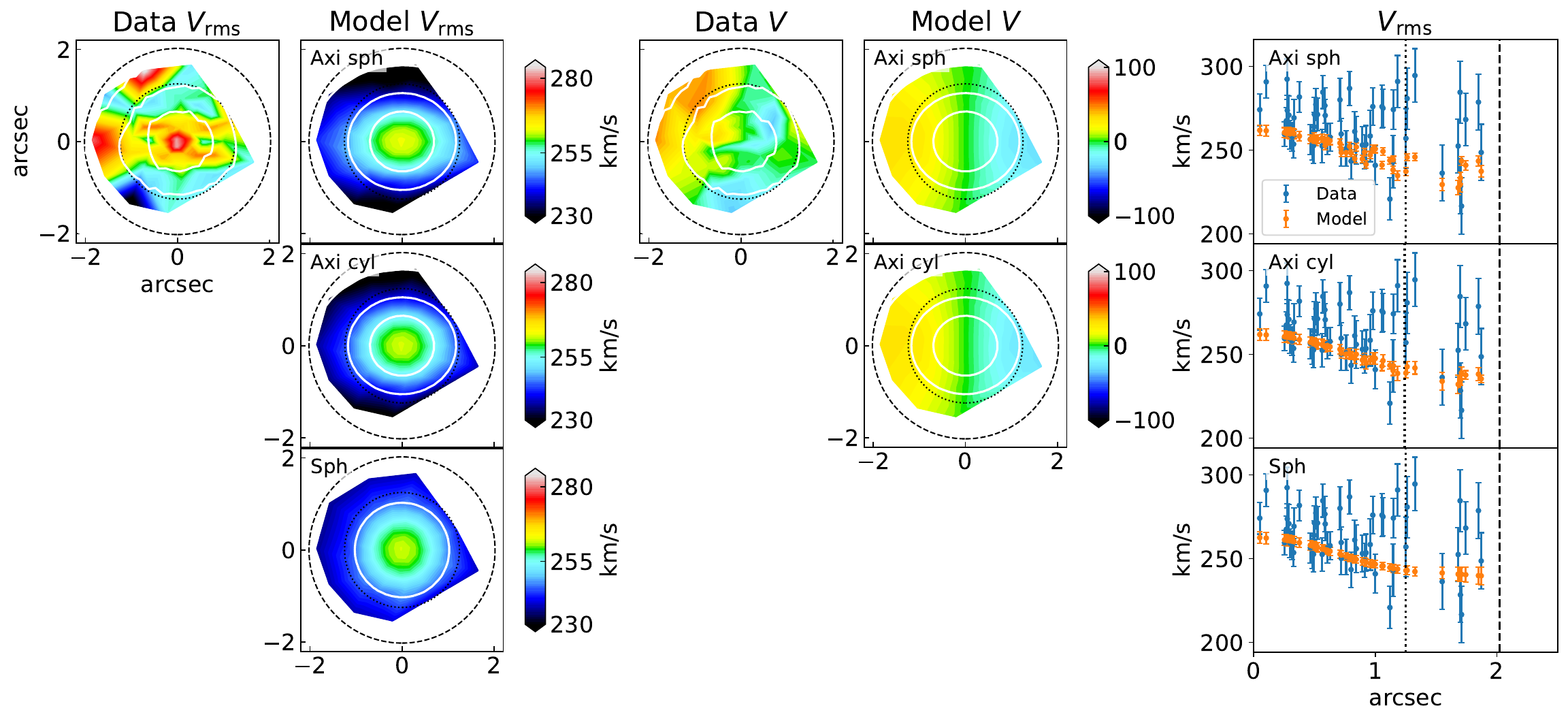}
    \includegraphics[width=0.85\linewidth]{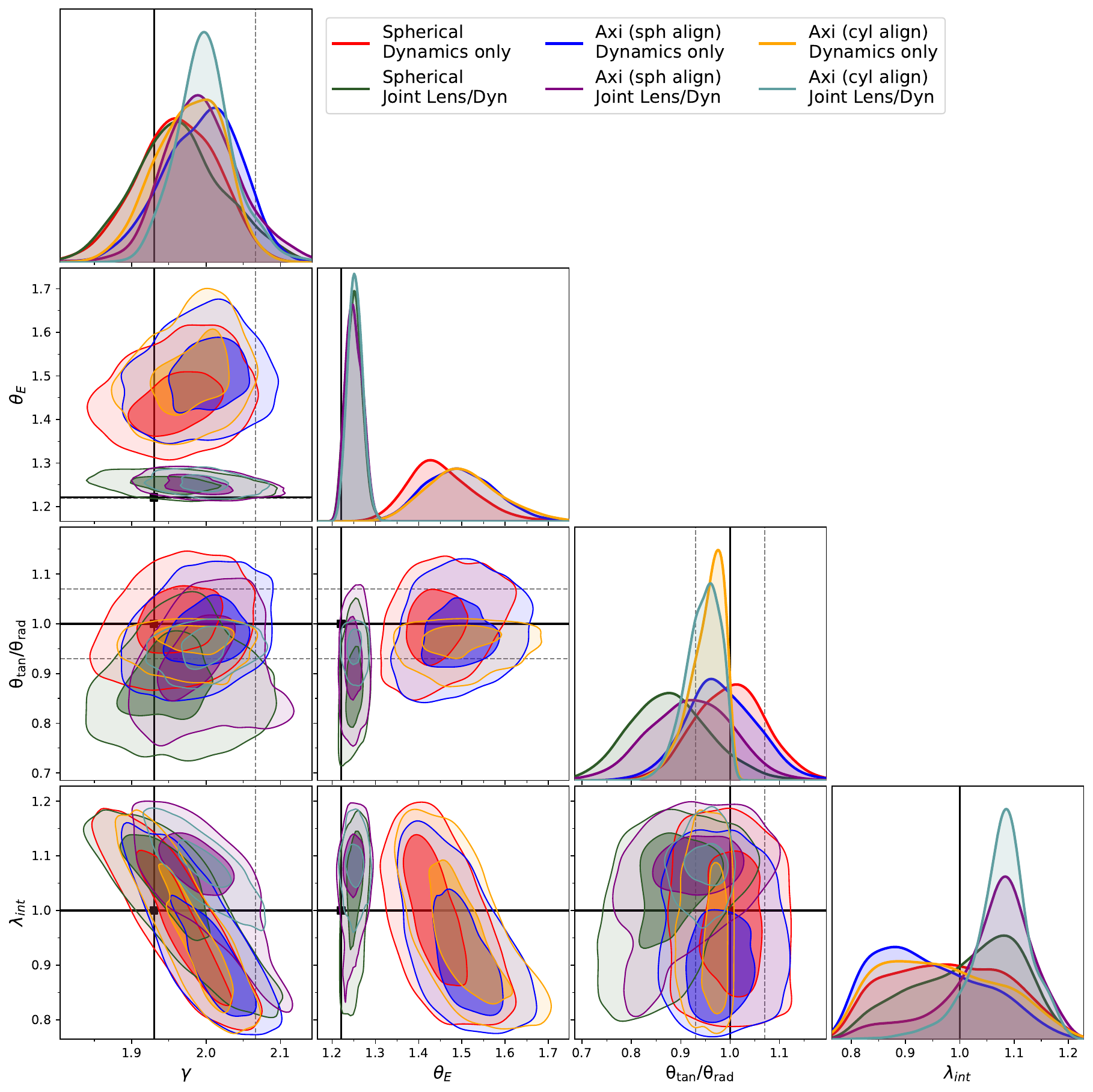}
    \caption{SDSSJ1627-0053}
    \label{fig:j1627}
\end{figure*}

\begin{figure*}
    \centering
    \includegraphics[width=\linewidth]{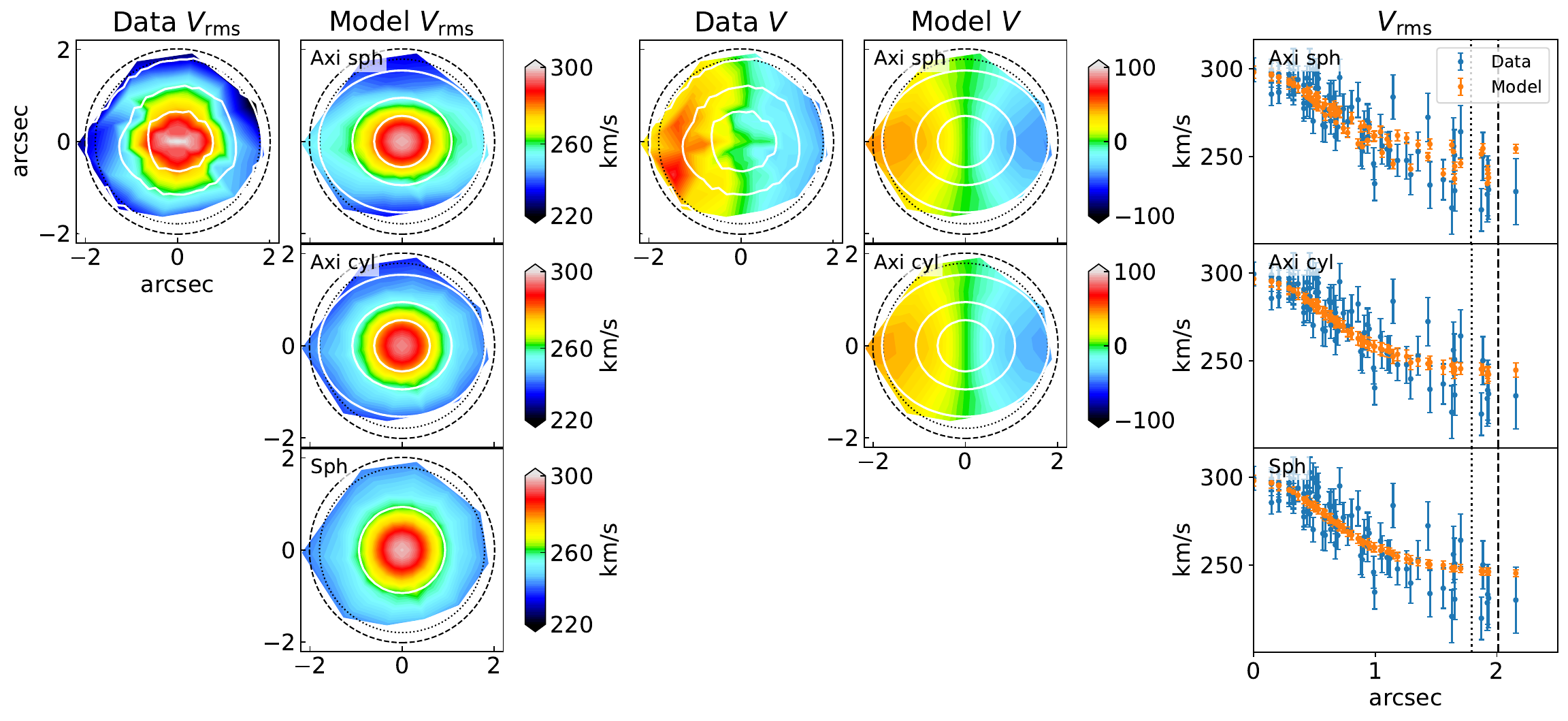}
    \includegraphics[width=0.85\linewidth]{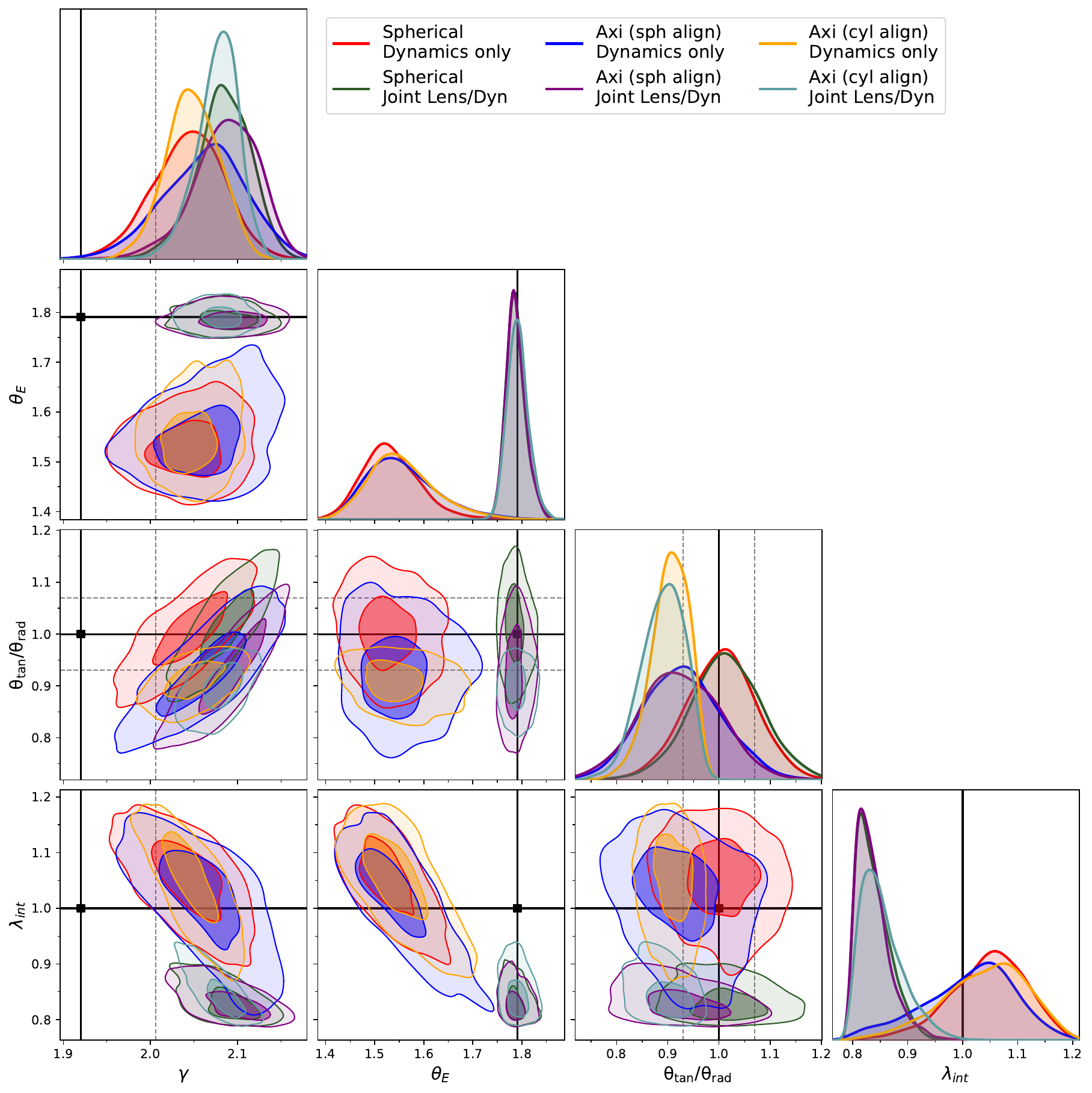}
    \caption{SDSSJ1630+4520}
    \label{fig:j1630}
\end{figure*}

\begin{figure*}
    \centering
    \includegraphics[width=\linewidth]{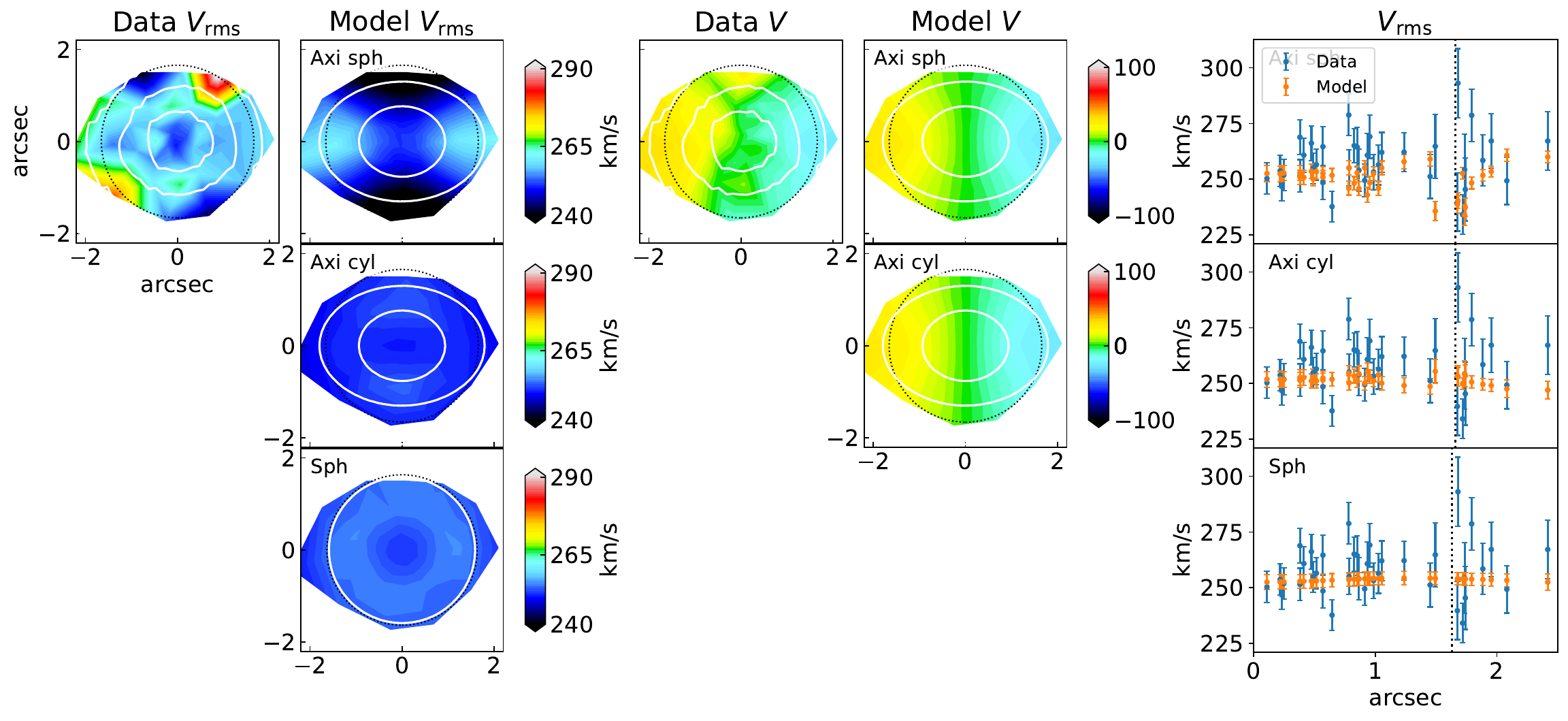}
    \includegraphics[width=0.85\linewidth]{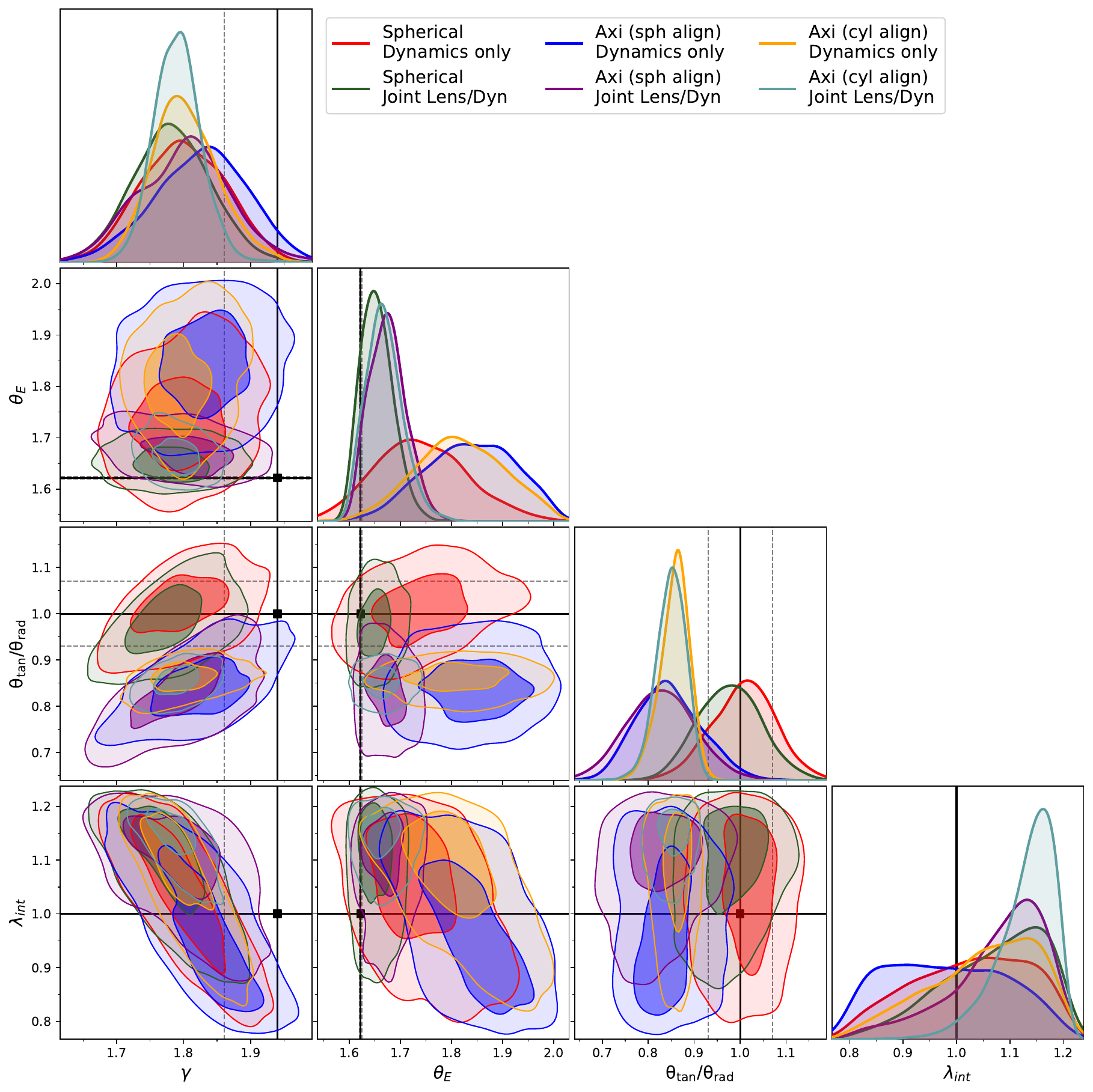}
    \caption{SDSSJ2303+1422}
    \label{fig:j2303}
\end{figure*}

In Figures \ref{fig:j0029}-\ref{fig:j2303}, we show data extracted from the IFS datacubes, JAM model best fits for joint lensing/dynamical models, and corner plots for each of the SLACS/KCWI sample. We show all kinematic maps of the data in a symmetrized manner for the sake of visualization only; the actual fits are performed at the raw bin level. For each object, the first column shows the $V_\mathrm{rms}$, followed by the best fit models under spherically-aligned axisymmetric, cylindrically-aligned axisymmetric, and spherical geometries. The third and fourth columns show the same for the mean bin velocity, i.e., the rotation. The rightmost column shows the radial profile of the $V_\mathrm{rms}$ for an alternative view that more easily shows data and model uncertainties $\delta V_\mathrm{rms}$. Dotted curves and lines indicate the best fit $\theta_\mathrm{E}$, and dashed curves and lines indicate the circularized effective radius. All kinematic maps extend to or beyond $\theta_\mathrm{E}$, but only 8/14 reach the effective radius. 

The corner plots show the overlap of all six models (dynamics-only and joint lensing/dynamics) in the key parameters of interest: power-law slope $\gamma$, Einstein radius $\rm \theta_E$, anisotropy ratio $\rm \theta_{tan}/\theta_{rad}$, and MST parameter $\rm \lambda_{int}$. The lens model posteriors for $\gamma$ and $\theta_\mathrm{E}$, as well as the spherical anisotropy prior are shown as black lines with dashed lines at $1\sigma$ intervals for reference. We show a line at $\rm \lambda_{int}=1$ to illustrate the effect of the mass sheet. Degenerate parameters are clearly expressed in the banana shape of the posteriors, particularly for the dynamics-only models. The $\theta_\mathrm{tan}/\theta_\mathrm{rad}$-$\gamma$ panel (middle row, left column) shows the mass-anisotropy degeneracy, while the $\lambda_\mathrm{int}$-$\theta_\mathrm{E}$ panel (bottom row, middle column) shows the mass-sheet degeneracy. In many cases, $\rm \lambda_{int}$ is poorly constrained for the dynamics-only models, but the strong constraints on $\theta_\mathrm{E}$ clearly constrain the parameter space allowed for $\lambda_\mathrm{int}$. The joint models only slightly improve constraints on the anisotropy beyond what the spatially resolved kinematics achieve on their own, which is primarily driven by the prior. 

\end{document}